\documentclass[]{mn2e}
\usepackage{epsfig}
\def\gsim{\vcenter{\hbox{$>$}\offinterlineskip\hbox{$\sim$}}}
\def\lsim{\vcenter{\hbox{$<$}\offinterlineskip\hbox{$\sim$}}}
\title[M\,33 monitoring. II]{The UK Infrared Telescope M\,33 monitoring
project. II. The star formation history in the central square kiloparsec}
\author[Javadi, van Loon \& Mirtorabi]{Atefeh Javadi$^{1,2}$, Jacco Th.\ van
Loon$^{2}$ and Mohammad Taghi Mirtorabi$^{3}$\\
$^{1}$School of Astronomy, Institute for Research in Fundamental Sciences
      (IPM), P.O.\ Box 19395-5531, Tehran, Iran\\
$^{2}$Astrophysics Group, Lennard-Jones Laboratories, Keele University,
      Staffordshire ST5 5BG, UK\\
$^{3}$Physics Department, Alzahra University, Vanak, Tehran, Iran}
\date{Resubmitted in February 2011}
\pagerange{\pageref{firstpage}--\pageref{lastpage}}
\pubyear{2011}
\begin{document}
\maketitle
\label{firstpage}
\begin{abstract}
We have conducted a near-infrared monitoring campaign at the UK InfraRed
Telescope (UKIRT), of the Local Group spiral galaxy M\,33 (Triangulum). The
main aim was to identify stars in the very final stage of their evolution, and
for which the luminosity is more directly related to the birth mass than the
more numerous less-evolved giant stars that continue to increase in
luminosity. In this second paper of the series, we construct the birth mass
function and hence derive the star formation history. The star formation rate
has varied between $\sim0.002$ and 0.007 M$_\odot$ yr$^{-1}$ kpc$^{-2}$. We
give evidence of two epochs of a star formation rate enhanced by a factor of a
few -- one that happened $\geq6$ Gyr ago and produced $\geq80$\% of the total
mass in stars, and one around 250 Myr ago that lasted $\sim200$ Myr and formed
$\leq6$\% of the mass in stars. We construct radial and azimuthal
distributions in the image plane and in the galaxy plane for populations
associated with old first-ascent red giant branch (RGB) stars,
intermediate-age Asymptotic Giant Branch (AGB) stars and young (massive) blue
and red supergiants. We find that the RGB stars follow a spheroidal
distribution, while younger stars follow a flat-disc distribution. The
intermediate-age population displays signs of a pseudo-bulge or possibly a
bar. The inner spiral arm pattern as recorded in mid-$19^{\rm th}$-century
drawings is confirmed. We interpret our findings as evidence for an old,
pressure-supported component and a younger disc formed 6 Gyr ago, with an
accretion event occurring 250 Myr ago giving rise to the compact nucleus in
M\,33. Our study provides support for recent Padova stellar evolution models
except that super-AGB stars likely reach low temperatures and thus high
mass-loss rates, supporting the super-AGB nature of the progenitors of
dust-enshrouded supernovae such as SN\,2008S.
\end{abstract}

\section{Introduction}

Spiral galaxies are among the more massive galaxies in the Universe, that in
the $\Lambda$-Cold Dark Matter paradigm are believed to have formed from the
agglomeration of smaller building blocks. They have gaseous, rotating discs,
where spiral arms and rotational shear set the conditions within which star
formation continues at a moderate rate in present times. In the central
regions of spiral galaxies, more ellipsoidal, largely stellar and
pressure-supported components exist -- nuclear star clusters, bulges, and
``bars'', but the relative importance of these components varies among
galaxies of this Hubble class. What determines the presence or absence of
these structures, and indeed the spiral arm pattern itself, is not understood.

The Local Group galaxy Triangulum (Hodierna 1654) -- hereafter referred to as
M\,33 (Messier 1771) -- offers us a unique opportunity to study a spiral
galaxy up close, and in particular to learn more about the structure and
evolution of the central regions of such galaxy, that in our own Milky Way are
heavily obscured by the intervening dusty Disc (van Loon et al.\ 2003;
Benjamin et al.\ 2005). Our viewing angle with respect to the M\,33 disc is
more favourable (56--$57^\circ$ -- Zaritsky, Elston \& Hill 1989; Deul \& van
der Hulst 1987) than that of the larger M\,31 (Andromeda), whilst the distance
to M\,33 is not much different from that to M\,31 ($\mu=24.9$ mag -- Bonanos
et al.\ 2006). The stellar content of M\,33 was summarised beautifully in the
review by van den Bergh (1991). The present-day star formation rate in the
central regions of M\,33 has been measured on the basis of optical photometry
of massive main-sequence stars and H$\alpha$ luminosity to be $\sim0.005$--0.1
M$_\odot$ yr$^{-1}$ kpc$^{-2}$ (Wilson, Scoville \& Rice 1991). Minniti,
Olszewski \& Rieke (1993) deduced from their infrared (IR) observations (in
the H-band at 1.6 $\mu$m) that a distinct star formation episode must have
occurred in the ``bulge'' of M\,33 within the last Gyr, unrelated to activity
within the disc -- they also showed that the bulge comprises also
intermediate-age stars in contrast to the old, metal-poor halo of M\,33. The
structure of the central region was investigated further at IR wavelengths by
Regan \& Vogel (1994) and Stephens \& Frogel (2002); the central nucleus had
been described in detail also by Lauer et al.\ (1998).

Galactic evolution is driven at the end-points of stellar evolution, where
copious mass loss returns chemically-enriched and sometimes dusty matter back
to the interstellar medium (ISM); the stellar winds of evolved stars and the
violent deaths of the most massive stars also inject energy and momentum into
the ISM, generating turbulence and galactic fountains when superbubbles pop as
they reach the ``surface'' of the galactic disc. The evolved stars are also
excellent tracers, not just of the feedback processes, but also of the
underlying populations, that were formed from millions to billions of years
prior to their appearance. The evolved phases of evolution generally represent
the most luminous, and often the coolest, making evolved stars brilliant
beacons at IR wavelengths, where it is also easier to see them deep inside
galaxies as dust is more tranpsarent at those longer wavelengths than in the
optical and ultraviolet where their main-sequence progenitors shine. The final
stages of stellar evolution of stars with main-sequence masses up to $M\sim30$
M$_\odot$ -- Asymptotic Giant Branch (AGB) stars and red supergiants -- are
characterised by strong radial pulsations of the cool atmospheric layers,
rendering them identifiable as long-period variables (LPVs) in photometric
monitoring campaigns spanning months to years (e.g., Whitelock, Feast \&
Catchpole 1991; Wood 2000; Ita et al.\ 2004a,b).

The main objectives of the project are: to construct the mass function of LPVs
and derive from this the star formation history in M\,33; to correlate spatial
distributions of the LPVs of different mass with galactic structures
(spheroid, disc and spiral arm components); to measure the rate at which dust
is produced and fed into the ISM; to establish correlations between the dust
production rate, luminosity, and amplitude of an LPV; and to compare the {\it
in situ} dust replenishment with the amount of pre-existing dust (see Javadi,
van Loon \& Mirtorabi 2011b). This is Paper II in the series, describing the
galactic structure and star formation history in the inner square kpc. Paper I
in the series presented the photometric catalogue of stars in the inner square
kpc (Javadi, van Loon and Mirtorabi 2011a). Subsequent papers in the series
will discuss the mass-loss mechanism and dust production rate (Paper III), and
the extension to a nearly square degree area covering much of the M\,33
optical disc (Paper IV).

\section{The input data and models}

In this section we briefly summarise the salient features of the catalogue of
variable stars that we produced (Paper I), and of the stellar evolution models
that we use here.

\subsection{The catalogue of variable stars that we use}

In Paper I we described the search for large-amplitude, long-period variable
stars in the central $4^\prime\times4^\prime$ (square kpc) of M\,33, with the
UIST imager on the United Kingdom IR Telescope (UKIRT) on Mauna Kea, Hawai'i.
The monitoring was done predominantly in the K-band (around a wavelength of
$\lambda=2.20$ $\mu$m), with additional observations in the J- and H-bands
($\lambda=1.28$ and 1.65 $\mu$m, respectively) mostly for the purpose of
obtaining colour information. Observations were carried out if the near-IR
stellar images were $<0.8^{\prime\prime}$. Each quadrant of the field was
observed at least 11 times spread over the period 2003--2007, with stars in
overlap regions being observed more often.

%
%
\begin{figure}
\centerline{\psfig{figure=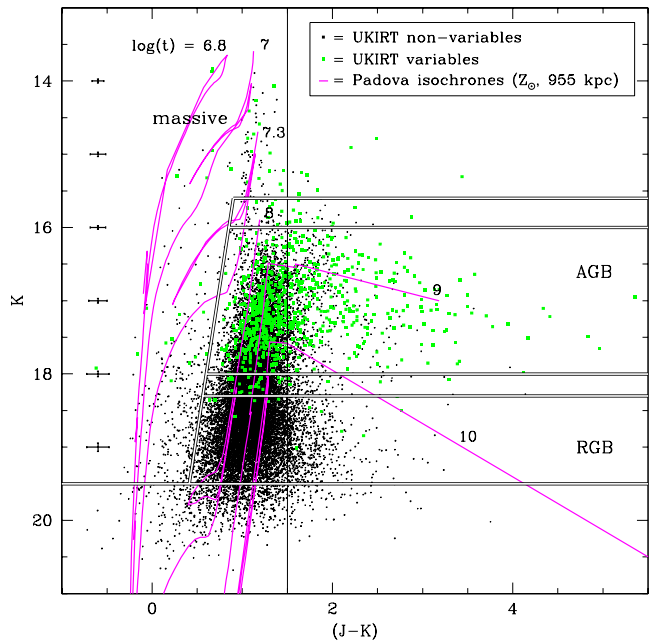,width=85mm}}
\caption[]{Colour--magnitude diagram of our UKIRT catalogue; variable stars
are plotted in green. Representative photometric errorbars are plotted.
Overlain are isochrones (Marigo et al.\ 2008), labelled by their logarithmic
ages -- these include circumstellar reddening, showing up as large excursions
towards red colours; carbonaceous dust (e.g., at $\log t=9$) and oxygeneous
dust (e.g., at $\log t=10$) have different reddening slopes. The vertical line
at $J-K=1.5$ mag indicates the colour criterion redwards of which a correction
is applied for this reddening. The thick black-and-white lines demarcate the
bulk of the populations of massive, AGB and RGB stars (separated by small
buffers in K-band magnitude to avoid cross-contamination -- see Section 4.2).}
\end{figure}

Photometry was obtained via fitting of the empirically-derived Point Spread
Function to all stars detected above a threshold in terms of the local sky
background noise. Hence a catalogue of 18,398 stars was produced. Fig.\ 1
shows these stars in a colour--magnitude diagram. The catalogue reaches a
completeness level of $>90$\% above the RGB tip around $K\sim18$ mag. For this
bright magnitude regime, the photometric accuracy is $<10$\%, and blending is
a concern in only $\sim1$\% of cases. Contamination of the sample by stars in
the Galactic foreground was demonstrated to be negligible.

The variability search was performed using the {\sc newtrial} routine (Stetson
1993) on the basis of the ``$L$'' variability index (Stetson 1996), which
includes an assessment of the variations in the measurements in comparison to
the measurement uncertainties, both in terms of their ratio and distribution.
Hence 812 variables were found, with estimated K-band amplitudes in the range
$A_{\rm K}\sim0.2$--2 mag (typically $A_{\rm K}\sim0.6$ mag). Their
distribution over magnitude and colour is shown in Fig.\ 1 (in green).

\subsection{The stellar evolution models that we use}

The best available theoretical models for our purpose are currently provided
by the Padova group (Marigo et al.\ 2008), for the following reasons:
\begin{itemize}
\item[$\bullet$]{They span an adequate range in birth mass ($0.8<M<30$
M$_\odot$), combining their own models for intermediate-mass stars ($M<7$
M$_\odot$) with Padova models for more massive stars ($M>7$ M$_\odot$;
Bertelli et al.\ 1994);}
\item[$\bullet$]{They are computed all the way through the thermal pulsing AGB
until they enter the post-AGB phase, in a manner that is consistent with the
computation of the preceding evolutionary stages. Crucially, it includes the
third dredge-up mixing of the stellar mantle as a result of the helium-burning
pulses, as well as the enhanced luminosity of massive AGB stars undergoing Hot
Bottom Burning (HBB -- Iben \& Renzini 1983);}
\item[$\bullet$]{They include molecular opacities that are important for the
cool atmospheres of red giants, allowing to describe the transformation from
oxygen-dominated M-type AGB stars to carbon stars in the approximate
birth-mass range $M\sim1.5$--4 M$_\odot$ (cf.\ Girardi \& Marigo 2007);}
\item[$\bullet$]{They include predictions for dust production in the winds of
LPVs and the associated reddening;}
\item[$\bullet$]{They include predictions for the radial pulsation;}
\item[$\bullet$]{They have been carefully transformed to various optical and
IR photometric systems;}
\item[$\bullet$]{They are available via a user-friendly internet interface.}
\end{itemize}
We encourage other groups to apply the same principles to their models, so
that their fundamental differences can be explored empirically to the greatest
effect. While we refer to Marigo et al.\ (2008) and Marigo \& Girardi (2007)
for the technical details pertaining to their models, and we have no way of
replacing them by different models, we point out a few essential aspects of
this set of Padova models below.

The Padova models do not include the effects of rotation or magnetic fields on
mixing within the stellar interior. Especially in massive stars these can have
important consequences for their luminosities and lifetimes, making them
brighter but live longer and evolve to cooler red supergiant stages (cf.\
Meynet \& Maeder 2000; Heger, Woosley \& Spruit 2005).

The Padova models employ convective overshoot (see Marigo \& Girardi 2007),
which prolongs the lifetime of a star. Hence stars of a given observed
luminosity will be linked to older ages and lower birth masses than in other,
classical models that do not allow for convective overshoot. This is mainly an
issue for intermediate-age populations.

The thermal pulsing AGB is not computed by solving the equations governing the
stellar internal structure, but in a synthetic way by adopting
parameterisations which include the efficiency and threshold temperature at
which third dredge-up sets in. These parameters have been calibrated
empirically against observations, for instance of the carbon star luminosity
function and star clusters in the Magellanic Clouds (Marigo \& Girardi 2007).
Consequently, it is possible that these calibrations are not a perfect match
to populations in galaxies other than the Magellanic Clouds. While stars in
M\,33 are not vastly different from those in the LMC, stars in the nuclear
region of M\,33 are probably more similar to solar-metallicity stars in the
Milky Way.

More critically important is to note that the published isochrones do {\it
not} include the excursions in luminosity and temperature resulting from the
thermal pulses, but only record the evolution through the hydrogen-shell
burning phases. While the latter dominates also on the TP-AGB, stars
undergoing the thermal-pulse excursions may erroneously be assigned to lower
or higher birth masses. This ultimately reduces the time resolution and
contrast of the derived star formation history somewhat.

The radial pulsation properties (period and mode) are computed as a function
of luminosity. Stars are expected to start pulsating in the first overtone as
soon as the thermal pulses start. The luminosity at which the transition to
the fundamental mode occurs is based upon the linear pulsation models from
Ostlie \& Cox (1986), while periods are computed from the mass and radius
based upon models from Fox \& Wood (1982) -- with some modifications to detail
as described in Marigo \& Girardi (2007; their Section 2.7) and Marigo et al.\
(2008; their Appendix). For the purpose of our work the main important feature
is the lifetime of the pulsation phase, not so much the mode or period. The
revisions made by Marigo et al.\ (2008) reduced the pulsation lifetime of the
most massive AGB stars by a factor of 2--5, which in our analysis would lead
to an increase in the associated star formation rate by the same amount (see
Section 3.2).

Spectral templates used to describe the photospheric emission of the coolest
giants come from empirical libraries for M-type stars (Fluks et al.\ 1994) or
synthetic spectra for carbon stars (Loidl, Lan\c{c}on \& J{\o}rgensen 2001).
These may not always be entirely appropriate, for instance at a different
metallicity, carbon-to-oxygen ratio or dynamical state of the pulsating
atmosphere. But the uncertainty in the bolometric corrections is modest in the
near-IR, likely $<0.1$ mag at K, and thus not a serious concern for our
project.

Mass loss truncates the evolution of AGB stars, depleting their mantles and
thereby avoiding the explosion of AGB stars of $M>1.4$ M$_\odot$. This becomes
important when the mass-loss rates exceed the nuclear burning rate (van Loon
et al.\ 1999). Mass loss also reduces the effective temperature and increases
the pulsation period, as the less-massive mantle inflates. Because all this
happens quite suddenly, it is as important to accurately predict {\it when}
this happens, as it is to accurately predict the {\it rate} of mass loss that
is attained. The formalism adopted in the Padova models is based upon
parameterisation with respect to pulsation period (cf.\ Vassialidis \& Wood
1993). While this has observational support it does mean that the accuracy of
the predicted mass loss relies critically on the model's accuracy with which
it predicts the pulsation properties. Alternative prescriptions for mass
loss exist that are defined in terms of luminosity and temperature (van Loon
et al.\ 2005). Mass loss is not as dramatic an issue for massive stars, which
do explode before shedding their mantle through stellar mass loss, though mass
loss could have an effect on the time these stars spend as a red supergiant.

Mass loss from cool stars also affects the photometric appearance of the star,
as dust forming in the wind gives rise to selective extinction and thus
diminishing brightness and reddening at optical/near-IR wavelengths. The
effect at near-IR wavelengths only becomes noticeable in the case of very
dense circumstellar envelopes. The Padova models incorporate these effects in
a two-step procedure: (1) predict the radial density of the dust envelope, and
(2) compute the resulting extinction of starlight at short wavelengths, and
re-emission by the grains at long wavelengths. The first step is based mainly
on the predictions from Ferrarotti \& Gail (2006) whilst the second step
follows standard radiative transfer computations as in Groenewegen (2006). The
adopted wind structure and dust content have observational support (e.g.,
Marshall et al.\ 2004) but there can be many reasons for deviations from this
standard, simplistic picture. Likewise, the optical properties of the grains
that are adopted in the radiative transfer computations are reasonable
(typical astronomical silicates or aluminium oxides for M-type stars and a
mixture of amorphous carbon and silicate carbide for carbon stars) but not
without uncertainties. For the purpose of the analysis presented here, the
most important uncertainty is in the direction of the reddening vector in the
near-IR colour--magnitude diagram. This is explored in some detail in Sections
3.1.1 and 4.1.1.

An important remaining uncertainty in the Padova models concerns the evolution
of super-AGB stars, with birth-masses $M\sim5$--10 M$_\odot$ (Siess 2007),
which in their models are not computed all the way through the thermal pulsing
phase and thus appear to terminate their evolution prematurely. We explore the
consequences in this paper and offer some guidance to the models.

Fig.\ 1 includes several representative isochrones, which show good
correspondence to the observed sequences of stars. The main regions in this
diagram where massive stars, AGB stars and RGB stars are found are demarcated
(see Section 4.2). Red AGB stars or massive stars, at $J-K>1.5$ mag, are
almost exclusively LPVs affected by {\it circumstellar} reddening; {\it
interstellar} reddening was proven to be unimportant (see Paper I). The
reddened stars form a minority among the sample of LPVs, but it is encouraging
to see that the isochrones cover the regime of reddened LPVs fairly well.

\section{From the brightnesses of variable stars to a star formation history}

The star formation history (SFH) is described by the star formation rate
(SFR), $\xi$, as a function of lapsed time, $t$. This function quantifies how
many solar masses of gas are converted into stars per year, and we here
express it per unit area projected on the sky. The combined mass of stars
created between times $t$ and $t+{\rm d}t$ is:
\begin{equation}
{\rm d}M(t)=\xi(t)\,{\rm d}t.
\end{equation}
This corresponds to a number, $N$, of stars formed:
\begin{equation}
{\rm d}N(t)={\rm d}M(t)\frac{\int_{\rm min}^{\rm max}f_{\rm IMF}(m)\,{\rm
d}m}{\int_{\rm min}^{\rm max}f_{\rm IMF}(m)m\,{\rm d}m},
\end{equation}
where $f_{\rm IMF}$ is the initial mass function, defined by:
\begin{equation}
f_{\rm IMF}=Am^{-\alpha},
\end{equation}
where $A$ is the normalization constant and $\alpha$ depends on the mass
range, following Kroupa (2001):
\begin{equation}
\alpha=\left\{
\begin{array}{lll}
+0.3\pm0.7 & {\rm for} & {\rm min}\leq m/{\rm M}_\odot<0.08 \\
+1.3\pm0.5 & {\rm for} & 0.08\leq m/{\rm M}_\odot<     0.50 \\
+2.3\pm0.3 & {\rm for} & 0.50\leq m/{\rm M}_\odot<{\rm max} \\
\end{array}
\right.
\end{equation}
The minimum and maximum of the stellar mass range are adopted to be 0.02 and
200 M$_\odot$, respectively. Reasonable changes in these values will result in
changes in the star formation rate by hardly more than a factor two.

The question is how many of these stars, $n$, are variable stars around time
$t$ which is when we observe them. Suppose stars with mass between $m(t)$ and
$m(t+{\rm d}t)$ satisfy this condition, then the number of variable stars
created between times $t$ and $t+{\rm d}t$ is:
\begin{equation}
{\rm d}n(t)={\rm d}N(t)\frac{\int_{m(t)}^{m(t+dt)}f_{\rm IMF}(m)\,{\rm
d}m}{\int_{\rm min}^{\rm max}f_{\rm IMF}(m)\,{\rm d}m}.
\end{equation}
Combining the above equations we have:
\begin{equation}
{\rm d}n(t)=\xi(t){\rm d}t\ \frac{\int_{m(t)}^{m(t+{\rm d}t)}f_{\rm
IMF}(m){\rm d}m}{\int_{\rm min}^{\rm max}f_{\rm IMF}(m)m\,{\rm d}m}.
\end{equation}

When we determine $\xi(t)$ over an age bin ${\rm d}t$, the number of variable
stars observed in that age bin, ${\rm d}n^\prime$, depends on the duration of
variability, $\delta t$, so:
\begin{equation}
{\rm d}n^\prime(t)={\rm d}n(t)\frac{\delta t}{{\rm d}t}.
\end{equation}
Inverting the above equations we obtain a relation between the observed
number of variable stars in a certain age bin, and the star formation rate
that long ago:
\begin{equation}
\xi(t)=\frac{{\rm d}n^\prime(t)}{\delta t}\ \frac{\int_{\rm min}^{\rm
max}f_{\rm IMF}(m)m\,{\rm d}m}{\int_{m(t)}^{m(t+{\rm d}t)}f_{\rm IMF}(m)\,{\rm
d}m}.
\end{equation}
Note that the value of the normalization constant $A$ in Eq.\ (3) does not
matter here.

\subsection{The relation between birth mass and K-band magnitude of
large-amplitude variable stars}

LPVs, which dominate the large-amplitude variables identified in our IR
monitoring programme, have reached the very final stages of their evolution,
and their brightness can thus be translated into their mass at birth by
employing theoretical evolutionary tracks, or in this case just as well by
isochrones. Inspection of the isochrones confirms that stars are at their
brightest in the K-band ($\lambda\sim2$ $\mu$m) when they also develop the
pulsations that give rise to their large-amplitude variability.

Hence we constructed the mass--luminosity relation for the K-band as depicted
in Fig.\ 2 (low- to intermediate-mass range) and Fig.\ 3 (which includes the
most massive stars), for four different values of the overall metallicity,
from super-solar -- as in massive elliptical galaxies and sub-populations in
the bulges of massive spiral galaxies such as the Milky Way -- to sub-solar
values appropriate for the Large and Small Magellanic Clouds. There are small
but noticeable differences. The central region of M\,33 is characterised by
approximately solar-metallicity ISM (Rosolowsky \& Simon 2008; Magrini et al.\
2009) and young and intermediate-age stellar populations (Massey et al.\ 1996;
Minniti et al.\ 1993), but also given that older stars will tend to have lower
metallicities we err on the side of caution and adopt the $Z=0.015$ result,
which is either consistent with solar metallicity or slightly sub-solar
metallicity depending on the exact metallicity adopted for the Sun ($Z=0.015$
is close to the value for the Sun determined in recent years -- cf.\ Asplund
et al.\ 2009).

The mass--luminosity relations show small excursions in places, e.g., for
$16<K<16.4$ mag at $Z=0.019$, which could be related to differences in the
bolometric corrections of M-type and carbon stars. However, some inaccuracies
result from limited accuracy in the models as well as the time-sampling of the
evolutionary tracks and isochrones, and where there is doubt we interpolate
over these small excursions by means of a spline fit, rendering a well-behaved
function.


%
%
\begin{figure}
\centerline{\psfig{figure=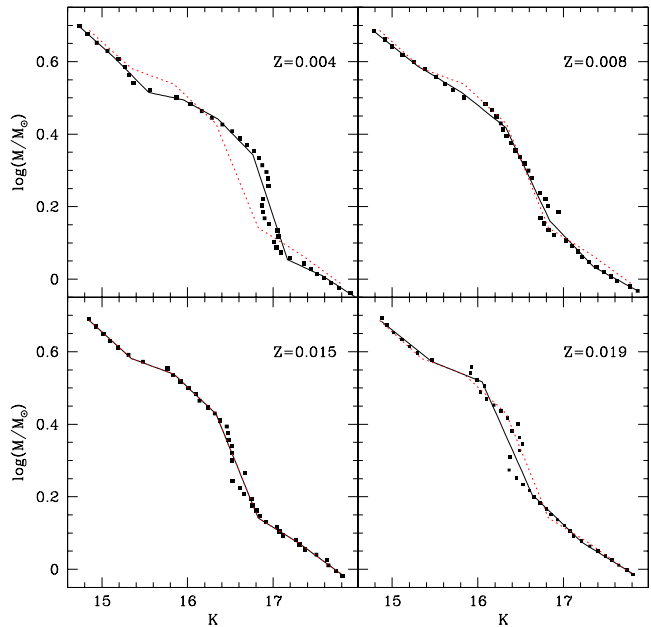,width=85mm}}
\caption[]{Mass--Luminosity relation for $Z=0.019$, 0.015, 0.008 and 0.004,
for the K-band magnitude range corresponding to the low- to intermediate-mass
range. Solid lines are the best linear spline fits, while the dotted lines are
the fit for $Z=0.015$ -- which we apply for the variables stars of M33 galaxy
-- for comparison.}
\end{figure}

%
%
\begin{figure}
\centerline{\psfig{figure=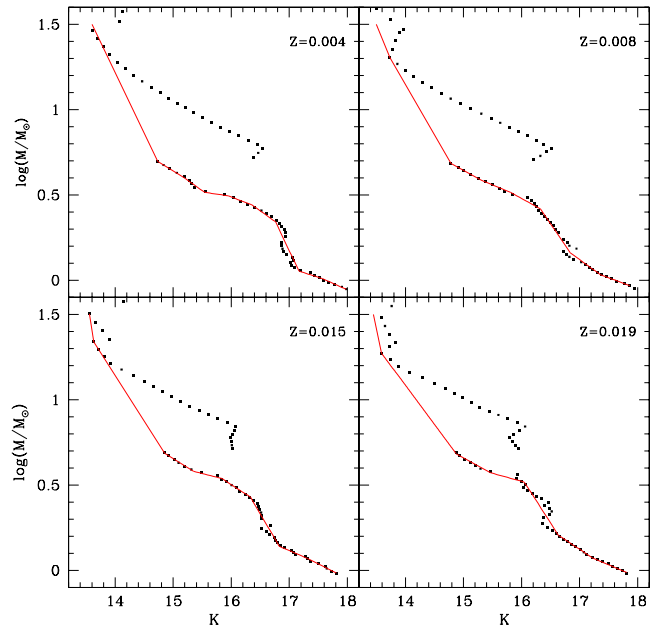,width=85mm}}
\caption[]{Mass--Luminosity relation for $Z=0.019$, 0.015, 0.008 and 0.004.
This graph shows the relation for the full range of K-band magnitudes (dots).
The solid lines are the linear spline fits, for the case in which the function
is interpolated across the super-AGB phase to massive red supergiants, i.e.\
for $0.7<\log(M/{\rm M}_\odot)<1.2$--1.3.}
\end{figure}

For $0.7<\log(M/{\rm M}_\odot)<1.2$--1.3, the isochrones suggest an excursion
towards fainter K-band magnitudes (Fig.\ 3), but as noted earlier the models
were not computed through to the end of the evolution of these super-AGB stars
(and possibly stars at the lower end of the mass range for red supergiants).
In our analysis, we follow two different approaches, and investigate the
impact on the derived star formation history:
\begin{itemize}
\item[1.]{We interpolate the Mass--Luminosity realtion (and the
Mass--Pulsation duration relation, see further) over this mass range to
connect the final luminosities of AGB stars to those of massive red
supergiants in a monotonic manner. Stars with $\log(M/{\rm M}_\odot)>1.5$ do
not appear to become red supergiants and they are therefore not expected to
exhibit periodic large-amplitude variability. This is our preferred solution,
which we vindicate in Section 4.1.5;}
\item[2.]{We accept the jump to fainter magnitudes from AGB to super-AGB
stars. This is an extreme measure, as we believe super-AGB stars should reach
high luminosities as they undergo HBB, though more massive red supergiants do
not. We postpone following this choice to Section 4.1.5, where we show that it
leads to physically unrealistic results.}
\end{itemize}
The coefficients of the best fitting function are obtained by minimisation of
$\chi^2$ with the Iraf task {\sc nfit1d} except where it was deemed more
appropriate to be guided by two end-points of isochrones, in which case no
estimates of uncertainty are attached. For case (1) the function is described
in Table 1.

%
%
\begin{table}
\caption[]{Mass--Luminosity relation, $\log{M\,{\rm [M}_\odot{\rm ]}}=aK+b$,
where $K$ is the K-band magnitude on the 2MASS system, for four values of
metallicity, for the case in which the function is interpolated across the
super-AGB phase to massive red supergiants.}
\begin{tabular}{ccc}
\hline\hline
\multicolumn{3}{c}{$Z=0.019$} \\
\hline
$a$              & $b$             & validity range       \\
\hline
$-1.627$         & \llap{2}3.3812  &        $K\leq13.590$ \\
$-0.458$         &         6.5002  & $13.590<K\leq14.875$ \\
$-0.188\pm0.053$ & $3.492\pm0.081$ & $14.875<K\leq15.462$ \\
$-0.094\pm0.045$ & $2.027\pm0.071$ & $15.462<K\leq16.049$ \\
$-0.532\pm0.033$ & $9.049\pm0.053$ & $16.049<K\leq16.635$ \\
$-0.225\pm0.029$ & $3.951\pm0.045$ & $16.635<K\leq17.222$ \\
$-0.143\pm0.041$ & $2.541\pm0.061$ &        $K>17.222$    \\
\hline\hline
\multicolumn{3}{c}{$Z=0.015$} \\
\hline
$a$              & $b$             & validity range       \\
\hline
$-2.198$         & \llap{3}1.3051  &        $K\leq13.633$ \\
$-0.535$         &         8.6253  & $13.633<K\leq14.854$ \\
$-0.213\pm0.027$ & $3.844\pm0.040$ & $14.854<K\leq15.347$ \\
$-0.088\pm0.018$ & $1.924\pm0.023$ & $15.347<K\leq15.840$ \\
$-0.222\pm0.021$ & $4.056\pm0.024$ & $15.840<K\leq16.332$ \\
$-0.582\pm0.012$ & $9.937\pm0.014$ & $16.332<K\leq16.825$ \\
$-0.147\pm0.013$ & $2.620\pm0.016$ & $16.825<K\leq17.318$ \\
$-0.173\pm0.022$ & $3.060\pm0.032$ &        $K>17.318$    \\
\hline\hline
\multicolumn{3}{c}{$Z=0.008$} \\
\hline
$a$              & $b$             & validity range       \\
\hline
$-0.840$         & \llap{1}2.840   &        $K\leq13.732$ \\
$-0.589$         &         9.391   & $13.732<K\leq14.787$ \\
$-0.188\pm0.028$ & $3.465\pm0.043$ & $14.787<K\leq15.300$ \\
$-0.142\pm0.027$ & $2.758\pm0.041$ & $15.300<K\leq15.813$ \\
$-0.188\pm0.022$ & $3.487\pm0.036$ & $15.813<K\leq16.327$ \\
$-0.501\pm0.016$ & $8.183\pm0.024$ & $16.327<K\leq16.840$ \\
$-0.248\pm0.018$ & $4.335\pm0.026$ & $16.840<K\leq17.353$ \\
$-0.128\pm0.025$ & $2.257\pm0.035$ &        $K>17.353$    \\
\hline\hline
\multicolumn{3}{c}{$Z=0.004$} \\
\hline
$a$              & $b$                     & validity range       \\
\hline
$-0.708$         &  \llap{1}1.1268         &        $K\leq14.734$ \\
$-0.209\pm0.051$ &         $3.783\pm0.077$ & $14.734<K\leq15.140$ \\
$-0.240\pm0.054$ &         $4.244\pm0.078$ & $15.140<K\leq15.545$ \\
$-0.050\pm0.060$ &         $0.565\pm0.091$ & $15.545<K\leq15.951$ \\
$-0.131\pm0.054$ &         $2.583\pm0.083$ & $15.951<K\leq16.356$ \\
$-0.243\pm0.041$ &         $4.409\pm0.057$ & $16.356<K\leq16.762$ \\
$-0.714\pm0.039$ & \llap{1}$2.310\pm0.055$ & $16.762<K\leq17.167$ \\
$-0.109\pm0.043$ &         $1.932\pm0.070$ & $17.167<K\leq17.573$ \\
$-0.153\pm0.046$ &         $2.690\pm0.067$ &        $K>17.573$    \\
\hline
\end{tabular}
\end{table}

\subsubsection{Correction for extinction by dust}

By determining the Mass--Luminosity relation we now can convert the K-band
magnitudes of the variable stars to their masses, and derive the present-day
mass function of variable stars. However, the K-band magnitude of a star seen
through a high dust column density is diminished, and should be corrected to
the value on the Mass--Luminosity relation. Although the isochrones of Marigo
et al.\ (2008) include the effects of circumstellar dust formation, these are
very uncertain and we prefer a simple de-reddening procedure. This would also
naturally account for any additional interstellar extinction, at least to some
extent\footnote{Interstellar dust will have somewhat different optical
properties from those of circumstellar dust. Then again, interstellar
extinction is not generally observed to be a major factor in this region of
M\,33, whereas circumstellar dust is evident in a considerable fraction of the
highly-evolved AGB stars -- see Paper I.}. The de-reddening correction is
related to the colour and magnitude of a star. In Fig.\ 1 we show the Marigo
et al.\ (2008) isochrones for a range of ages. It immediately becomes clear
that the reddening for carbon stars (e.g., at $t=1$ Gyr i.e.\ $\log t=9$)
differs from that of oxygen-rich stars (e.g., at $t=10$ Gyr i.e.\ $\log
t=10$), the former more quickly reddening compared to the extinction in the
K-band. The average slope of the reddening of carbon stars is 0.52 mag of
K-band extinction per magnitude of $J-K$ reddening, whilst it is 0.72 mag
mag$^{-1}$ for oxygen-rich stars. The locus to which we translate the reddened
star is at $J-K=1.25$ mag, but we apply the correction to those variable stars
that have $J-K>1.5$ mag to allow some deviations from the unreddened locus as
a result of photometric inaccuracies and variability. Hence the correction
equation is:
\begin{equation}
K_0=K+a(1.25-(J-K))
\end{equation}
If only K- and H-band data are available but no J-band data then we apply the
following correction derived from the K versus H--K diagram:
\begin{equation}
K_0=K+a(0.3-(H-K))
\end{equation}
In the case where only K-band but no J- nor H-band data are available, we
adopt $J-K=4$ mag (roughyl equivalent to where we start losing stars too red
for our survey depth), and use Eq.\ (9). The constants in these equations are
$a=1.06$ for carbon stars and $a=1.6$ for oxygen-rich stars\footnote{In the
remainder we assume all photometry has been corrected for extinction, and we
do not explicitly use the subscript ``0''.}.

It is thus important to know which stars have carbonaceous dust and which ones
have oxygeneous dust. As described in Paper I, the only catalogue available
that uses an identification technique based on specific spectral features is
the one published by Rowe et al.\ (2005). Then, following the recommendation
by those auhtors, we set limits on the photometric errors in their catalogue
of $<0.05$ mag, resulting in just three variable carbon stars. This is
insufficient to decide which red stars in our catalogue are likely to be
carbon stars. Setting no limit on errors resulted in five variable carbon
stars, of which three have $J-K>1.5$ mag. However, we expect from theory and
observations in the LMC (Groenewegen \& de Jong 1993; van Loon, Marshall \&
Zijlstra 2005; Girardi \& Marigo 2007) that solar and slightly sub-solar
metallicity carbon stars arise from stars in a birth mass range of
$\approx1.5$--4 M$_\odot$: $3^{\rm rd}$ dredge-up is not strong enough to turn
lower-mass AGB stars into carbon stars, while HBB prevents carbon to enrich
the surface of more massive AGB stars. Hence we first apply the carbon-dust
correction to all reddened stars -- if the derived mass is in the 1.5--4
M$_\odot$ range then we accept that star as a carbon star, else we apply the
oxygeneous dust correction.

In Fig.\ 4 we show the resulting PMF for the variable stars in the central
square kpc of M\,33. As indicated in Paper I, nine of the large-amplitude
variable stars show indications of blending so we removed those stars from our
anlysis and we derive the star formation history using the remaining 803
variables. We applied the de-reddening correction as described above, as well
as {\it only} a carbonaceous dust correction and {\it only} an oxygeneous dust
correction: the differences are slight and mostly reflected in an $\sin50$\%
scatter in the $\sim2$ M$_\odot$ region.

%
%
\begin{figure}
\centerline{\psfig{figure=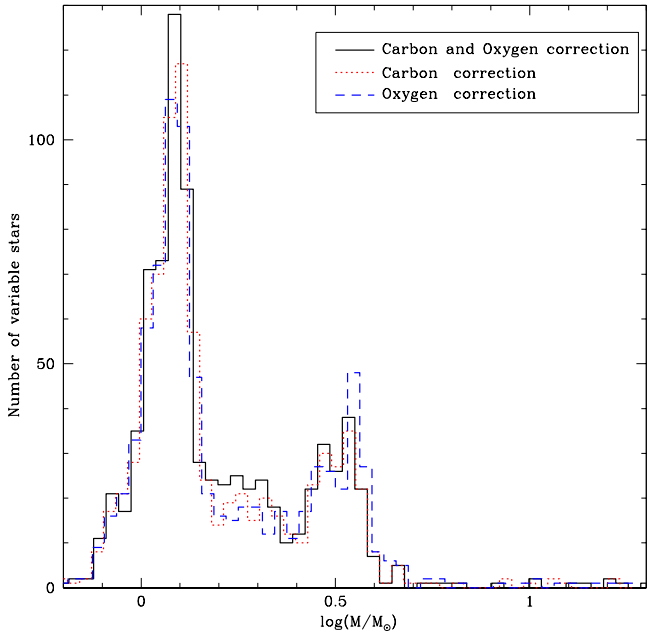,width=85mm}}
\caption[]{Present-day mass function of large-amplitude variable stars in the
central square kpc of M\,33. The solid black line is the histogram obtained by
applying the de-reddening correction dependent on the mass of the star, the
dotted red and dashed blue lines are the results obtained when applying {\it
only} the carbonaceous dust correction or {\it only} the oxygeneous dust
correction, respectively.}
\end{figure}

\subsection{The mass-dependent duration of large-amplitude variability}

Eq.\ (1) contains a correction factor for the duration in which a star is seen
to pulsate with large amplitude. For instance, massive stars evolve faster and
are captured in their pulsating phase only briefly, whereas lower-mass stars
spend longer in that phase and are thus more likely and more numerous to
appear in variability surveys -- aside from the effects of the IMF, and of the
SFH that we are after.

To derive this correction we followed the predictions from the Marigo et al.\
(2008) isochrones, so as to be consistent within our analysis. The isochrone
tables indicate the phase of strong radial pulsation, and we take this entire
duration as the time during which the star would have been identified in our
survey (Paper I) as a variable star. We derived a function of mass for this
duration of pulsation by fitting a multiple-Gaussian function to these
isochrone values. As we investigate two cases -- one in which we interpolate
between the most massive AGB stars and the more massive red supergiants, and
another in which we accept the regression in K-band brightness over the
super-AGB star mass regime -- we also take these two approaches with regard to
the pulsation-duration versus mass function (the Mass--Pulsation relation): if
the super-AGB stars {\it do} evolve to higher luminosities and cooler
temperatures then they will also be more likely to, and longer, pulsate.

Hence, for case (1) we fitted three Gaussian funtions to the Mass--Pulsation
diagram and interpolated over the same mass range as in the Mass--Luminosity
diagram. The result is tabulated in Table 2 and shown in Fig.\ 5 (top panel).
For case (2), we fit two independent sets of Gaussian functions -- one for the
low-mass stars and one for the high-mass stars (see Section 4.1.5).

%
%
\begin{table}
\caption[]{Mass--Pulsation relation, $\log(\delta t/t) = D + \sum_{i=1}^{3}a_i
\exp\left( (\log{M\,{\rm [M}_\odot{\rm ]}}-b_i)^2/(2c_i^2)\right)$, where
$\delta t$ is the pulsation duration and $t$ the age of the star, for four
values of metallicity, for the case in which the function is interpolated
across the super-AGB phase to massive red supergiants.}
\begin{tabular}{ccccc}
\hline\hline
\multicolumn{5}{c}{$Z=0.019$} \\
\hline
$D$     & $i$ & $a$  & $b$   & $c$ \\
\hline
$-4.42$ & 1 & 3.04 & 1.271 & 0.277 \\
        & 2 & 0.74 & 0.510 & 0.068 \\
        & 3 & 1.35 & 0.478 & 0.331 \\
\hline\hline
\multicolumn{5}{c}{$Z=0.015$} \\
\hline
$D$     & $i$ & $a$  & $b$   & $c$ \\
\hline
$-3.63$ & 1 & 2.18 & 1.238 & 0.238 \\
        & 2 & 1.36 & 0.514 & 0.127 \\
        & 3 & 0.30 & 0.228 & 0.084 \\
\hline\hline
\multicolumn{5}{c}{$Z=0.008$} \\
\hline
$D$     & $i$ & $a$  & $b$   & $c$ \\
\hline
$-3.96$ & 1 & 2.34 & 1.281 & 0.378 \\
        & 2 & 1.32 & 0.460 & 0.165 \\
        & 3 & 0.38 & 0.145 & 0.067 \\
\hline\hline
\multicolumn{5}{c}{$Z=0.004$} \\
\hline
$D$     & $i$ & $a$  & $b$   & $c$ \\
\hline
$-4.00$ & 1 & 2.29 & 1.217 & 0.408 \\
        & 2 & 0.84 & 0.524 & 0.093 \\
        & 3 & 0.87 & 0.206 & 0.088 \\
\hline
\end{tabular}
\end{table}

%
%
\begin{figure}
\centerline{\psfig{figure=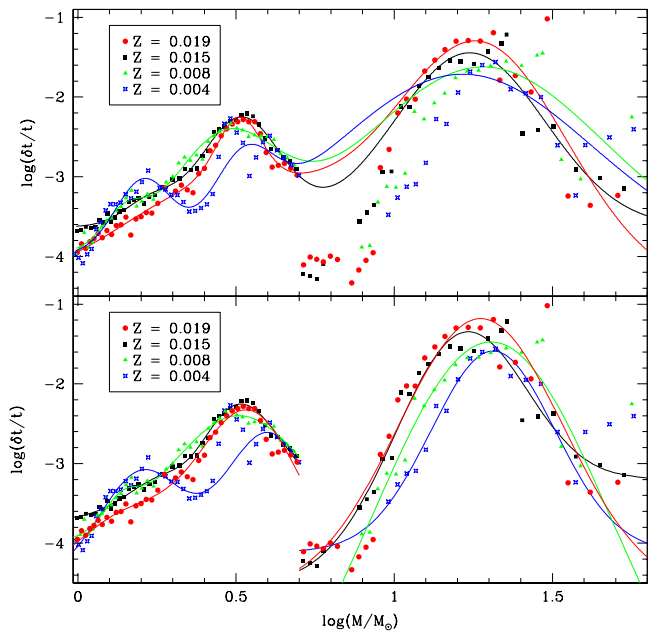,width=85mm}}
\caption[]{Mass--Pulsation relation. The points show the ratio of pulsation
duration to age as tabulated in the Marigo et al.\ (2008) isochrones; the
solid lines are our multiple-Gaussian fits to these data: interpolating over
the super-AGB regime (top panel, for four choices of metallicity $Z$) and
accepting the jump in K-band brightness between the AGB and super-AGB (bottom
panel).}
\end{figure}

\subsection{From a mass distribution to a star formation rate as a function of
time}

Here we have all the relations required to calculate the SFR, except that we
still need to relate the birth mass of a variable star we see now, to the time
lapsed since its formation. This transformation is easily derived from the
Marigo et al.\ (2008) isochrones. The Mass--Age relation is displayed in Fig.\
6, in which the data points are derived from the isochrones and the solid line
is a function we fit to the data (the coefficients of which are listed in
Table 3).

%
%
\begin{figure}
\centerline{\psfig{figure=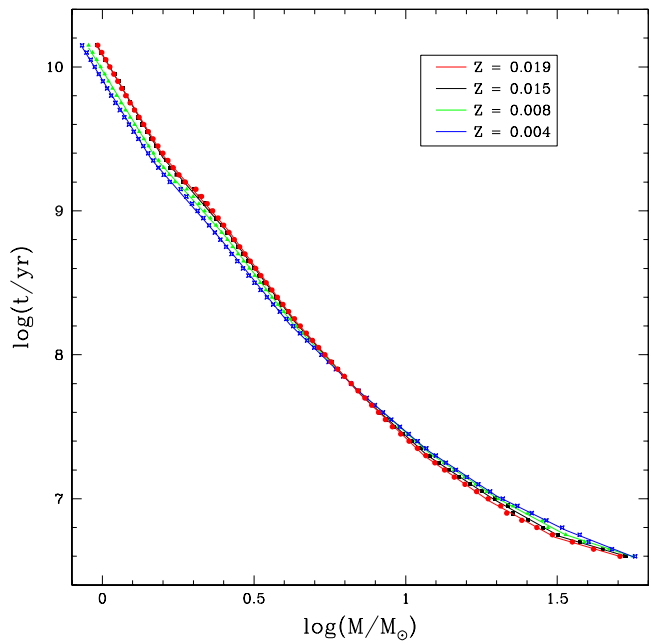,width=85mm}}
\caption[]{(Birth) Mass--Age relation for AGB stars and red supergiants
derived from the Marigo et al.\ (2008) isochrones for four choices of
metallicity (dots, $Z$), along with linear spline fits.}
\end{figure}

%
%
\begin{table}
\caption[]{Mass--Age relation, $\log{t\,{\rm [yr]}}=a\,\log{M\,{\rm
[M}_\odot{\rm ]}}+b$.}
\begin{tabular}{ccc}
\hline
\hline
\multicolumn{3}{c}{$Z=0.019$} \\
\hline
$a$              & $b$                     & validity range           \\
\hline
$-3.480\pm0.008$ & \llap{1}$0.082\pm0.019$ &       $\log{M}\leq0.200$ \\
$-2.433\pm0.007$ &         $9.873\pm0.010$ & $0.200<\log{M}\leq0.415$ \\
$-2.866\pm0.007$ & \llap{1}$0.054\pm0.010$ & $0.415<\log{M}\leq0.631$ \\
$-2.381\pm0.007$ &         $9.750\pm0.010$ & $0.631<\log{M}\leq0.846$ \\
$-2.027\pm0.008$ &         $9.450\pm0.012$ & $0.846<\log{M}\leq1.061$ \\
$-1.463\pm0.009$ &         $8.852\pm0.013$ & $1.061<\log{M}\leq1.277$ \\
$-1.168\pm0.011$ &         $8.474\pm0.016$ & $1.277<\log{M}\leq1.492$ \\
$-0.621\pm0.014$ &         $7.600\pm0.019$ &       $\log{M}>1.492$    \\
\hline\hline
\multicolumn{3}{c}{$Z=0.015$} \\
\hline
$a$              & $b$                     & validity range           \\
\hline
$-3.062\pm0.007$ & \llap{1}$0.096\pm0.011$ &       $\log{M}\leq0.237$ \\
$-2.425\pm0.007$ &         $9.945\pm0.010$ & $0.237<\log{M}\leq0.455$ \\
$-2.824\pm0.007$ & \llap{1}$0.127\pm0.010$ & $0.455<\log{M}\leq0.674$ \\
$-2.316\pm0.007$ &         $9.786\pm0.011$ & $0.674<\log{M}\leq0.892$ \\
$-1.926\pm0.008$ &         $9.438\pm0.012$ & $0.892<\log{M}\leq1.110$ \\
$-1.399\pm0.009$ &         $8.853\pm0.013$ & $1.110<\log{M}\leq1.328$ \\
$-1.180\pm0.063$ &         $8.562\pm0.067$ & $1.328<\log{M}\leq1.546$ \\
$-0.625\pm0.013$ &         $7.704\pm0.018$ &       $\log{M}>1.546$    \\
\hline\hline
\multicolumn{3}{c}{$Z=0.008$} \\
\hline
$a$              & $b$                     & validity range           \\
\hline
$-3.461\pm0.008$ &         $9.976\pm0.012$ &       $\log{M}\leq0.179$ \\
$-2.347\pm0.007$ &         $9.776\pm0.011$ & $0.179<\log{M}\leq0.404$ \\
$-2.727\pm0.008$ &         $9.930\pm0.011$ & $0.404<\log{M}\leq0.628$ \\
$-2.154\pm0.008$ &         $9.570\pm0.014$ & $0.628<\log{M}\leq0.853$ \\
$-1.848\pm0.009$ &         $9.309\pm0.015$ & $0.853<\log{M}\leq1.077$ \\
$-1.398\pm0.010$ &         $8.825\pm0.015$ & $1.077<\log{M}\leq1.302$ \\
$-1.134\pm0.012$ &         $8.451\pm0.017$ & $1.302<\log{M}\leq1.526$ \\
$-0.681\pm0.015$ &         $7.790\pm0.021$ &       $\log{M}>1.526$    \\
\hline\hline
\multicolumn{3}{c}{$Z=0.004$} \\
\hline
$a$              & $b$                     & validity range           \\
\hline
$-3.416\pm0.007$ &         $9.910\pm0.011$ &       $\log{M}\leq0.161$ \\
$-2.436\pm0.007$ &         $9.752\pm0.010$ & $0.161<\log{M}\leq0.389$ \\
$-2.588\pm0.007$ &         $9.812\pm0.011$ & $0.389<\log{M}\leq0.617$ \\
$-2.043\pm0.007$ &         $9.475\pm0.011$ & $0.617<\log{M}\leq0.845$ \\
$-1.812\pm0.008$ &         $9.280\pm0.012$ & $0.845<\log{M}\leq1.073$ \\
$-1.395\pm0.010$ &         $8.832\pm0.013$ & $1.073<\log{M}\leq1.301$ \\
$-1.036\pm0.011$ &         $8.365\pm0.016$ & $1.301<\log{M}\leq1.529$ \\
$-0.829\pm0.014$ &         $8.048\pm0.020$ &       $\log{M}>1.529$    \\
\hline
\end{tabular}
\end{table}

Now we have got all the tools ready to derive the SFH. For each star, first we
examine whether the star is reddened or not and if so, we add the de-reddening
correction using the mass-dependent coefficients. Then, by using the
coefficients from Table 1 for $Z=0.015$, we convert the K-band magnitude to
the birth mass and by using Table 2 we estimate the pulsation duration, the
inverse of which is the weight assigned to the star. We do this by
interpolation across the super-AGB regime -- later (in Section 4.1.5) we
investigate the difference obtained by following the isochrones exactly. Then,
we convert the birth mass to lapsed time (age). The only remaining correction
to be applied is the IMF.

A not unimportant detail related to the presentation of a SFH is the way it is
binned in age. The younger, massive variable stars are many times fewer than
the old, low-mass variable stars and inadequate binning can either lead to
spurious peaks in the SFR or mask any such real bursts. From a statistical
point of view there is advantage is assuring that each bin contains the same
number of stars, so the SFR values have uniform uncertainties. To accomplish
this, we first ordered the stars by mass, then start counting until we reach a
given number, at which moment we start counting stars for the subsequent bin.
Eventually, the IMF correction is applied based on the minimum and maximum
mass of each bin.

\section{The star formation history in M\,33}

The obtained SFH for the central square kpc of M\,33 is shown in Fig.\ 7. The
horizontal ``error bars'' correspond to the start and end time for which the
SFR was calculated in that bin. The oldest bin is only shown in part as it
extends to (unrealistically) large ages. The only reason it is plotted is to
show that, as expected, the star formation rate we estimate for ages exceeding
the Hubble time is negligible. We prefer not to rescale the diagrams so as not
to degrade their visibility in the ranges where it matters.

The SFH is characterised by a major epoch of formation $\approx4$--8 Gyr ago
($\log t=9.6$--9.9), peaking around 6 Gyr ago ($\log t=9.8$) at a level about
three times as high as during the subsequent couple of Gyr. This corroborates
the result obtained by Cioni et al.\ (2008) that the nuclear region of M\,33
is on average nearly as old as the outer disc, which they found to have an age
of $\sim6$ Gyr. The peak at 6 Gyr may not be real, though: the Mass--Pulsation
relation suggests that low-mass stars do not pulsate much. While this is only
true in a relative sense (as they live long), the exact level and duration of
pulsation for those stars may be somewhat uncertain. The SFR may thus well
have been as high, or higher earlier than 6 Gyr ago. We are confident, though,
that the SFR has dropped significantly since 5 Gyr ago ($\log t<9.7$).

%
%
\begin{figure}
\centerline{\psfig{figure=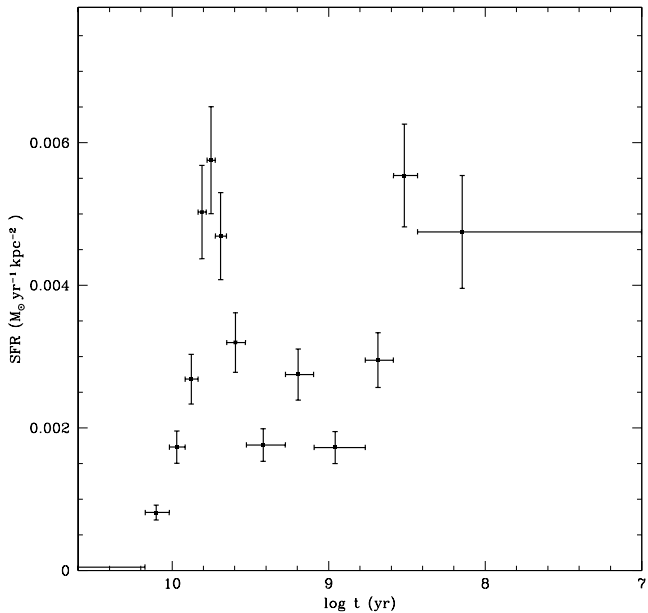,width=85mm}}
\caption[]{The star formation history in the central square kpc of M\,33
derived from near-IR AGB-star and red-supergiant variables.}
\end{figure}

A more recent epoch of enhanced star formation is seen too, occurring
$\sim200$--300 Myr ago ($\log t=8.3$--8.5), nearly reaching the same level as
the 6-Gyr peak ($\log t=9.8$). Since then the rate of star formation appears
to have diminished again. This would confirm the early recognition by Minniti
et al.\ (1993) of a substantial star formation episode in the ``bulge'' of
M\,33 within the past Gyr ($\log t<9$). It would be difficult to reliably
recover such short-duration enhancements in SFR at intermediate ages more than
a Gyr ago. For instance the higher bin around 1.5 Gyr ago ($\log t=9.2$) might
well correspond to such period of enhanced star formation, but on the basis of
the star number statistics (represented by the vertical errorbars in Fig.\ 7)
we must conclude that such assertion would be unfounded.

The present-day SFR in the inner disc of M\,33 (over a 4 kpc$^2$ area) has
been estimated by Wilson et al.\ (1991) on the basis of H$\alpha$ and far-IR
emission at $\xi=0.01$ M$_\odot$ yr$^{-1}$ kpc$^{-2}$. At
$\xi=0.0056\pm0.0008$ M$_\odot$ yr$^{-1}$ kpc$^{-2}$ both the ancient and
recent epochs of enhanced star formation that we have identified appear to be
similar in intensity to the estimates by Wilson et al. Theirs are valid for
the most recent few $10^7$ yr, however. In our sample of variable stars we
count six stars with ages $t<20$ Myr ($\log t<7.3$, birth masses $>12$
M$_\odot$). The SFR derived from this recently-formed population is
$\xi=0.007\pm0.003$ M$_\odot$ yr$^{-1}$ kpc$^{-2}$: while the errorbar on this
value is rather large the value is in excellent agreement with the value
derived by Wilson et al.\ on the basis of a very different method. This is
very reassuring, and to a certain degree validates our methodology.

Based on our SFH, the total mass in stars formed over the history of M\,33
amounts to $3.9\times10^7$ M$_\odot$ kpc$^{-2}$, of which $3.1\times10^7$
M$_\odot$ kpc$^{-2}$ (or 80\%) was formed at ages $t>4$ Gyr ($\log t>9.6$). In
contrast, the most recent epoch of enhanced star formation, at $t<500$ Myr
($\log t<8.7$), has only produced $2.4\times10^6$ M$_\odot$ kpc$^{-2}$ (or
6\%). These estimates assume that there has not been a net migration into or
out of the central square kpc region that we have monitored. Especially for
the older stars this may not be true, and they could have preferentially
migrated out of the central region as a result of dynamical relaxation (cf.\
van Loon et al.\ 2003). So the star formation rate -- and total stellar mass
formed -- that we derive for ages $t\gsim 5$ Gyr ($\log t\gsim 9.7$) may have
been underestimated.

Long, Charles \& Dubus (2002) determined that the nucleus (central few arcsec)
of M\,33 underwent two discrete epochs of star formation, one a Gyr ago ($\log
t=9$) which formed 4\% of the $2\times10^6$ M$_\odot$ in stellar mass in the
nuclear star cluster (cf.\ Kormendy \& McClure 1993), and another only
$\sim40$ Myr ago ($\log t\sim7.6$) which added just half a per cent to the
cluster's mass. Curiously, while timed differently (though with a degree of
uncertainty), this resembles the increment in stellar mass (6\%) that we found
to have formed within the past 0.5 Gyr ($\log t<8.7$) in the surrounding
central disc. At the star formation rate that we derived for the past 20 Myr
($\log t<7.3$), the most recent of the two star formation events in the
nucleus would have lasted approximately one Myr ($\log \Delta t=6$), which is
a typical timescale for star cluster formation. These comparisons are
suggestive of some connection between the nucleus and surrounding disc, and
possibly of a more similar timing of events.

\subsection{Variation in assumptions}

In this section we examine how the derived SFH changes if we change our
assumptions: the selection or omission of reddened stars, the amplitude
threshold, the distance modulus, and the treatment of the super-AGB stars.

\subsubsection{Reddened stars}

We can examine the contribution of the reddened stars to the SFH that we
derive from the large-amplitude variables. Our catalogue of variable stars in
the central square kpc of M\,33 contains 318 stars that have $J-K>1.5$ mag
(but one of these is removed because it was affected by blending), which
corresponds to the significant fraction of 39\% of all variable stars. These
stars need a correction to their K-band magnitudes. Much redder stars, though
rarer, may have been missed entirely by less IR-sensitive surveys. Of all
reddened stars, 45\% were identified as variable. Of the remaining 55\%, some
might well also be variable stars not identified as such in our survey, though
it is likely that some fraction of this concerns non-variable stars seen
behind interstellar dust clouds.

Fig.\ 8 shows the SFH we derive based solely on reddened ($J-K>1.5$ mag)
variable stars identified in our survey. As before, a mass-dependent
de-reddening correction has been applied to account for a mixture of carbon
and oxygen-rich stars. Two things stand out immediately: the major historical
star formation epoch at 4--10 Gyr ($\log t=9.6$--10) disappeared and also the
SFR in the most recent 250 Myr ($\log t<8.4$) has dropped. On the other hand,
reddened stars dominate the variable stars associated with the SFR derived at
intermediate ages, 0.25--2.5 Gyr ago ($\log t<8.4$--9.4). Most of these are
expected to be carbon stars. Clearly, massive and low-mass stars do not
feature prominently among the reddened stars. We argue, as we did in Paper I,
that the low-mass stars may become very red (and thus be missed even by our IR
survey) but possibly only very late in their pulsating phase, whilst massive
stars do not envelope themselves in optically-thick dust envelopes except in
the most extreme, and thus rare cases (cf.\ van Loon et al.\ 1997). In
conclusion, to obtain a complete SFH based on variable stars one must account
for reddened stars so as to not underestimate the SFR at intermediate ages,
while the earliest and most recent SFR is determined by optically bright(er)
variable stars.

%
%
\begin{figure}
\centerline{\psfig{figure=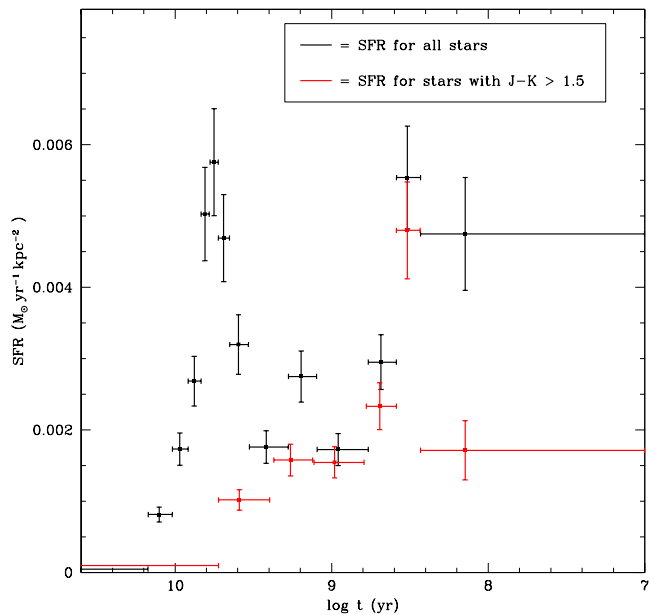,width=85mm}}
\caption[]{Contribution to the derived star formation history by reddened
variable stars with $J-K>1.5$ mag (in red), compared to the original result
(in black).}
\end{figure}

Our decision to only correct for reddening stars with $J-K>1.5$ mag, and to
correct them to a locus of $J-K=1.25$ mag, may introduce some uncertainty. If
either choice were made differently, the corrected brightness and hence
inferred birth mass and thus age would be different. We had introduced the
0.25 mag ``buffer'' between the locus of unreddened stars and the threshold
for correcting for reddening, to allow for photometric uncertainty which, if
corrected as if it were due to reddening would lead to systematic deviations.
The photometric uncertainty in the colour is typically $\sim0.1$ mag (Fig.\
1). The amplitudes of variability are typically 0.6 mag; while the colours
vary less, with only a few measurements in the bands other than the K-band the
time-average colours may be inaccurate by a few tenths of a magnitude. This,
and the morphology in the colour--magnitude diagram (Fig.\ 1), supports a
$J-K>1.5$ mag threshold. One could argue for a shift of this, or of the
unreddened locus, by 0.1 mag or so, but this would only lead to a shift
(largely systematic) in unreddened K-band brightness by a similar magnitude.
This would resemble varying the distance modulus by such amount. In Section
4.1.3 we investigate such shift but by as much as 0.4 mag. The effect of a 0.1
mag adjustment appears to be of little significance.

\subsubsection{Amplitude threshold}

How sensitive is the derived SFH to the exact threshold of our survey to the
detection of variability? To gain some idea of this effect, we set a selection
threshold on the estimated K-band amplitude (cf.\ Paper I) of $A_{\rm K}>0.5$
mag. This leaves us with 581 variable stars (excluding 9 stars affected by
blending -- see Paper I), still nearly 3/4 of the original selection. It is
likely that more stars were removed that have either low or high mass, as the
former may not pulsate as vigorously (McDonald et al.\ 2010) while the latter
do but not in terms of magnitude -- which is a relative quantity, in relation
to their high luminosity (cf.\ van Loon et al.\ 2008).

Fig.\ 9 shows the resulting SFH. The overall trend is little affected. Apart
from an overall modest drop in the SFR, the effect seems to affect mostly the
massive stars formed in the past 400 Myr ($\log t<8.6$). The smaller
amplitudes of low-mass stars is not apparent, but our SFH also does not extend
into the past as far as the ages of old Galactic globular clusters ($t>10$ Gyr
i.e.\ $\log t>10$) upon which the evidence for diminished pulsation in
low-mass red giants is largely based.

%
%
\begin{figure}
\centerline{\psfig{figure=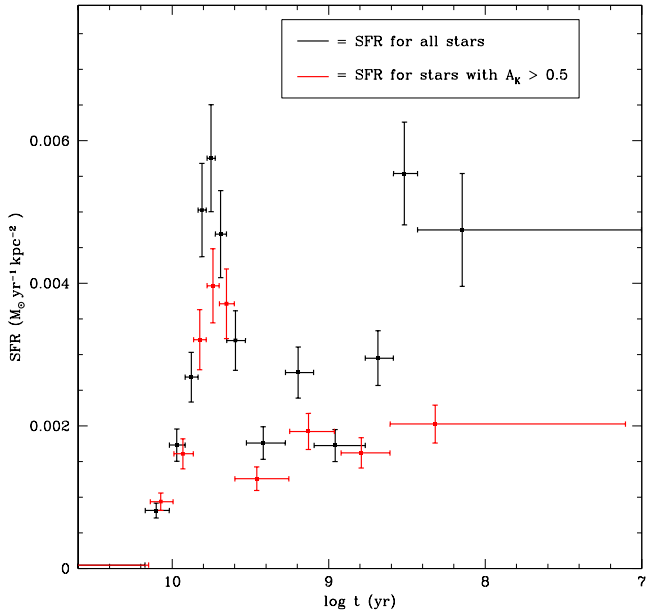,width=85mm}}
\caption[]{Star formation history derived after imposing a higher threshold on
the estimated amplitude of variability, $A_{\rm K}> 0.5$ mag (in red),
compared to the original result (in black).}
\end{figure}

\subsubsection{Distance modulus}

In our analysis we adopted a distance modulus of $\mu=24.9$ mag, which is
found by several authors via different methods including eclipsing binaries,
and which was found to match the theoretical isochrones quite well to our IR
colour--magnitude diagrams. However, some authors have found lower values for
the distance modulus, around $\mu=24.5$ mag (Scowcroft et al.\ 2009). Changing
the distance modulus will change the absolute magnitude of stars and hence
also the mass of stars, but the Mass--Pulsation relation will remain unchanged
as it does not depend on distance. This means that changing the distance
modulus will affect the SFH we derive.

Fig.\ 10 shows the SFH for $\mu=24.9$ (as before) and $\mu=24.5$ mag. Overall
the SFH has a similar behaviour despite the 20\% difference in distance, with
a major ancient epoch of star formation and enhanced star formation in recent
times. However, not surprisingly, the SFH is shifted towards earlier epochs.
The SFR seems to be diminished at all later times; this is caused by the same
effect, stars having been shifted to earlier times, piling up at ages
$t\sim10$ Gyr ($\log t\sim10$).

%
%
\begin{figure}
\centerline{\psfig{figure=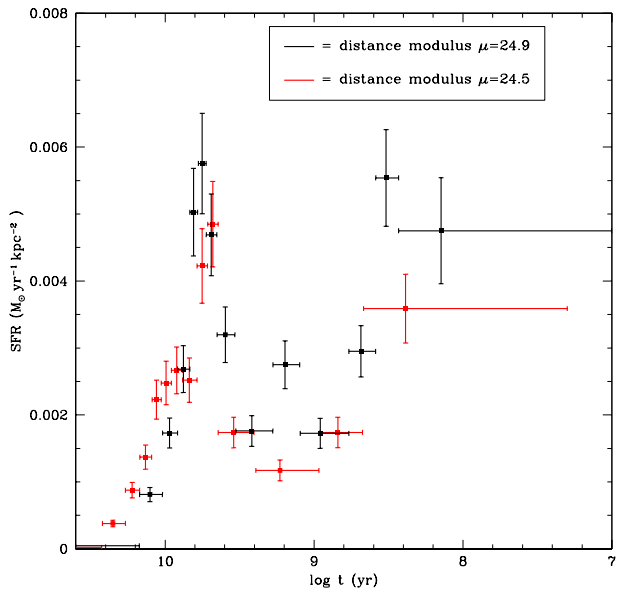,width=85mm}}
\caption[]{Star formation history compared for two choices of the distance
modulus: $\mu=24.9$ (black) and $\mu=24.5$ mag (red).}
\end{figure}

\subsubsection{Metallicity}

While the metallicity of the pulsating red giants in the centre of M\,33 is
quite reliably known, the oldest pulsating stars might be somewhat deficient
in metals compared to the young and intermediate-age $Z=0.015$ population that
formed after the initial period of star formation from which most chemical
enrichment resulted. Investigating the metallicity dependence of our results
also lays the foundation for the application of the method to the outskirts of
the M\,33 disc, or to altogether different galaxies.

%
%
\begin{figure}
\centerline{\psfig{figure=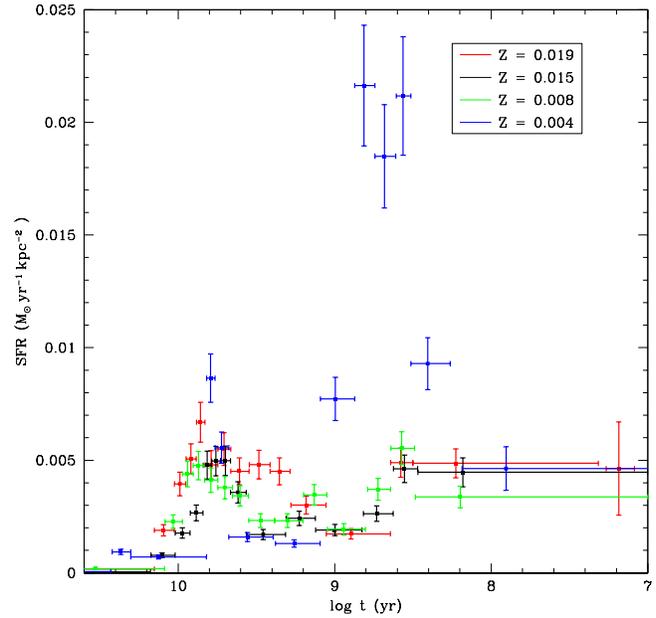,width=85mm}}
\caption[]{Star formation history compared for four choices of metallicity:
$Z=0.004$ (blue), $Z=0.008$ (green), $Z=0.015$ (black, appropriate for the
M\,33 core) and $Z=0.019$ (red).}
\end{figure}

The parameterisations of the Mass--Luminosity relation (Figs.\ 2 \& 3; Table
1), Mass--Pulsation relation (Fig.\ 5, top; Table 2) and Mass--Age relation
(Fig.\ 6; Table 3) have all been determined for four different values of the
metallicity: $Z=0.004$ (appropriate for the Small Magellanic Cloud), $Z=0.008$
(appropriate for the Large Magellanic Cloud), $Z=0.015$ (appropriate for the
centre of M\,33) and $Z=0.019$ (super-solar). The resulting star formation
histories are shown in Fig.\ 11.

The result from adopting $Z=0.008$ hardly differs from the one we obtained for
$Z=0.015$. Apart from pushing the early epoch of star formation back by one or
two Gyr there are no important deviations outside a $\sim20$\% margin.
Adopting $Z=0.019$ mainly extends the early epoch of star formation to as
recent as 2 Gyr ago ($\log t=9.3$) -- the star formation rate would become
fairly constant over the entire history of M\,33 except for a lull in activity
$\sim0.4$--2 Gyr ago ($\log t=8.6$--9.3). This strange result can be dismissed
as such high metallicity is an unrealistic choice for M\,33. Adopting
$Z=0.004$ yields a ``burst'' of star formation 0.2--1 Gyr ago ($\log
t=8.3$--9); whilst not impossible this is attributable to the shorter duration
of pulsation predicted by the models (Fig.\ 5) for the birth mass range of
2--4 M$_\odot$ ($\log M=0.3$--0.6; these are carbon stars) in combination with
the observed high numbers of pulsating stars in that mass range. This low
metallicity is not characteristic of the M\,33 population. In conclusion, the
results we have derived by adopting $Z=0.015$ are robust against reasonable
deviations from this value for the metallicity.

\subsubsection{Super-AGB stars}

Previously, we have interpolated the Mass--Luminosity and Mass--Pulsation
relations across the super-AGB and low-mass red supergiant regime, as the
theoretical models had not been computed until the very end of these stars'
evolution and we may thus have missed the coolest, strongly-pulsating phases
that these stars experience before ending their lives. Here, we examine how
our results change if we do take the endpoints of the theoretical isochrones
at face value.

Fig.\ 12 shows the revised Mass--Luminosity relation: the AGB portion, at
$\log({M/{\rm M}_\odot})<0.7$, remains unchanged but the description of the
more massive portion has been altered to follow the isochrones more closely
(Table 4). We still keep that portion in itself monotonic, for simplicity. The
small ranges where the slope has the opposite sign will not make a very
different contribution because of this. Clearly, however, stars of
$K\sim15^{\rm th}$ magnitude have two solutions for the mass, and we will need
to devise some procedure to distribute the stars over these two solutions.

%
%
\begin{figure}
\centerline{\psfig{figure=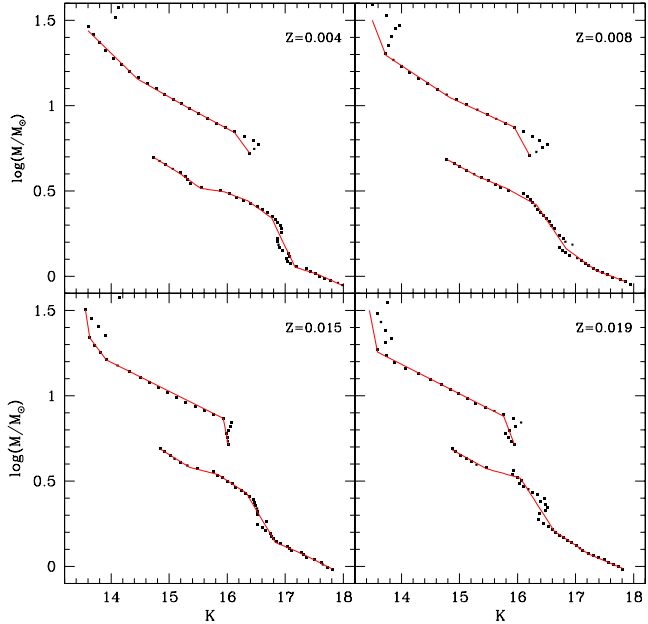,width=85mm}}
\caption[]{Mass--Luminosity relation as in Fig.\ 3 but with the revised
parametrisation for the super-AGB stars and low-mass red supergiants. The
solid lines are linear spline fits, requiring monotony everywhere except for
the discontinuity at $\log(M/{\rm M}_\odot)=0.7$.}
\end{figure}

%
%
\begin{table}
\caption[]{Mass--Luminosity relation, $\log{M\,{\rm [M}_\odot{\rm ]}}=aK+b$,
where $K$ is the K-band magnitude on the 2MASS system, for four values of
metallicity, for the case in which the endpoints of the isochrones covering
the super-AGB phase have been taken at face value. Only the high-mass portion
is given here -- the low-mass portion remains as in Table 1.}
\begin{tabular}{ccc}
\hline\hline
\multicolumn{3}{c}{$Z=0.019$} \\
\hline
$a$      & $b$           & validity range       \\
\hline
$-1.749$ & \llap{2}5.034 &        $K\leq13.590$ \\
$-0.173$ &         3.609 & $13.590<K\leq15.762$ \\
$-0.889$ & \llap{1}4.893 & $15.762<K\leq15.948$ \\
\hline\hline
\multicolumn{3}{c}{$Z=0.015$} \\
\hline
$a$      & $b$           & validity range       \\
\hline
$-2.198$ & \llap{3}1.305 &        $K\leq13.633$ \\
$-0.451$ &         7.491 & $13.633<K\leq13.917$ \\
$-0.170$ &         3.582 & $13.917<K\leq15.930$ \\
$-1.673$ & \llap{1}7.311 & $15.930<K\leq16.023$ \\
\hline\hline
\multicolumn{3}{c}{$Z=0.008$} \\
\hline
$a$              & $b$                     & validity range       \\
\hline
$-0.888$         &  \llap{1}3.486          &        $K\leq13.732$ \\
$-0.223\pm0.004$ &         $4.355\pm0.008$ & $13.732<K\leq14.837$ \\
$-0.157\pm0.006$ &         $3.378\pm0.009$ & $14.837<K\leq15.942$ \\
$-0.623\pm0.003$ & \llap{1}$0.811\pm0.006$ & $15.942<K\leq16.211$ \\
\hline\hline
\multicolumn{3}{c}{$Z=0.004$} \\
\hline
$a$              & $b$                     & validity range       \\
\hline
$-0.272\pm0.012$ &         $5.148\pm0.018$ &        $K\leq14.453$ \\
$-0.681\pm0.012$ & \llap{1}$1.005\pm0.018$ & $14.453<K\leq15.288$ \\
$-0.183\pm0.013$ &         $3.801\pm0.019$ & $15.288<K\leq16.124$ \\
$-0.458\pm0.007$ &         $8.222\pm0.014$ & $16.124<K\leq16.386$ \\
\hline
\end{tabular}
\end{table}

%
%
\begin{table}
\caption[]{Mass--Pulsation relation, $\log(\delta t/t) = D + \sum_{i=1}^{3}a_i
\exp\left( (\log{M\,{\rm [M}_\odot{\rm ]}}-b_i)^2/(2c_i^2)\right)$, where
$\delta t$ is the pulsation duration and $t$ the age of the star, for four
values of metallicity, for the case in which the endpoints of the isochrones
covering the super-AGB phase have been taken at face value.}
\begin{tabular}{cccccc}
\hline\hline
\multicolumn{6}{c}{$Z=0.019$} \\
\hline
$D$     & $i$ &          $a_i$ & $b_i$ & $c_i$ & validity range    \\
\hline
$-4.63$ & 1   &           3.45 & 1.273 & 0.262 & $\log{M}  > 0.68$ \\
        & 2   &           0.27 & 0.611 & 0.016 & $\log{M}  > 0.68$ \\
$-4.45$ & 1   &           2.10 & 0.538 & 0.166 & $\log{M}\leq0.68$ \\
        & 2   &           0.75 & 0.149 & 0.153 & $\log{M}\leq0.68$ \\
\hline\hline
\multicolumn{6}{c}{$Z=0.015$} \\
\hline
$D$     & $i$ &          $a_i$ & $b_i$ & $c_i$ & validity range    \\
\hline
$-4.52$ & 1   &           1.28 & 1.840 & 0.362 & $\log{M}  > 0.68$ \\
        & 2   &           2.88 & 1.211 & 0.215 & $\log{M}  > 0.68$ \\
$-3.72$ & 1   &           1.63 & 0.468 & 0.180 & $\log{M}\leq0.68$ \\
        & 2   & \llap{$-$}0.53 & 0.361 & 0.085 & $\log{M}\leq0.68$ \\
\hline\hline
\multicolumn{6}{c}{$Z=0.008$} \\
\hline
$D$     & $i$ &          $a_i$ & $b_i$ & $c_i$ & validity range    \\
\hline
$-7.13$ & 1   &           5.66 & 1.306 & 0.381 & $\log{M}  > 0.68$ \\
$-4.02$ & 1   &           1.62 & 0.509 & 0.217 & $\log{M}\leq0.68$ \\
        & 2   &           0.30 & 0.137 & 0.054 & $\log{M}\leq0.68$ \\ 
\hline\hline
\multicolumn{6}{c}{$Z=0.004$} \\
\hline
$D$     & $i$ &          $a_i$ & $b_i$ & $c_i$ & validity range    \\
\hline
$-4.11$ & 1   &           2.52 & 1.313 & 0.195 & $\log{M}  > 0.68$ \\
$-4.44$ & 1   &           1.82 & 0.602 & 0.142 & $\log{M}\leq0.68$ \\
        & 2   &           1.33 & 0.202 & 0.132 & $\log{M}\leq0.68$ \\
\hline
\end{tabular}
\end{table}

In the overlap region, we let us be guided by the IMF as to the relative
fractions of the star that are to be assigned to the low- and high-mass
solutions, respectively:
\begin{equation}
p_{\rm low}=\int_{m_1}^{m_2}f_{\rm IMF}\,{\rm d}m
\end{equation}
\begin{equation}
p_{\rm high}=\int_{m_1^\prime}^{m_2^\prime}f_{\rm IMF}\,{\rm d}m 
\end{equation} 
Here, $m_1$ and $m_2$ are the masses calculated using the low-mass solution
for each star, between $K-\delta K$ and $K+\delta K$, where $\delta K$ is the
error on the K-band magnitude; likewise, $m_1^\prime$ and $m_2^\prime$ delimit
the high-mass range. Hence we obtain the normalised probabilities for the low-
and high-mass solutions as follows:
\begin{equation}
P_{\rm low}=\frac{p_{\rm low}}{p_{\rm low}+p_{\rm high}}
\end{equation}
\begin{equation}
P_{\rm high}=\frac{p_{\rm high}}{p_{\rm low}+p_{\rm high}}
\end{equation}
so that each star still has unit weight, $P_{\rm low}+P_{\rm high}=1$.

The remaining procedure follows that described above, but using the modified
Mass--Pulsation relation given in Table 5 and displayed in Fig.\ 5. The
resulting SFH is shown in Fig.\ 13.

%
%
\begin{figure}
\centerline{\psfig{figure=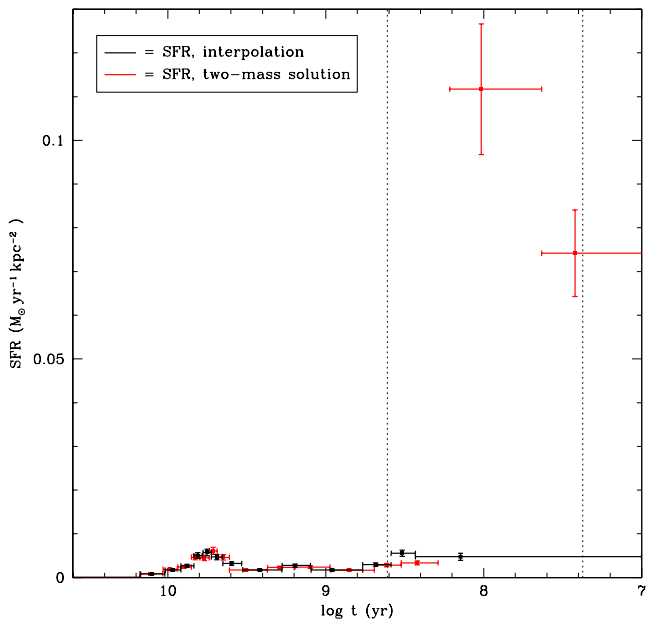,width=85mm}}
\caption[]{Star formation history for $Z=0.015$, derived by assigning stars to
dual solutions for their mass (in red), compared to the original result (in
black); the vertical dotted lines indicate the extent of the overlap region.}
\end{figure}

In the overlap region we notice a sudden increase in the SFR, to $\sim20$
times larger than the levels derived by interpolation. Such extreme behaviour
right where the modification is made is highly implausible. Could it result
from the implicit assumption that the SFR was the same for both solutions for
the star's mass, underlying the formulae for the fractions $p_{\rm low}$ and
$p_{\rm high}$? The inflated values for the SFR are due to the high-mass
branch of the dual solution. If the SFR in reality declined with time, then
smaller fractions of stars ought to be assigned to that solution and the SFR
would come down. However, the SFR must decline precipitously in order for the
high-mass branch to be depleted, essentially resulting in fewer than one star
to be observed in that regime. This would be very coincidental and thus also
highly unlikely. Rather, we find it more likely that the high values for the
SFR result from unrealistically-brief adopted durations for their variability,
$\delta t$ (cf.\ Fig.\ 5).

We note that the older of the three age bins in the overlap region (at $\log
t\sim8.4$) now has a somewhat reduced SFR: this is because these stars have
been partly assigned to the high-mass branch of the solution whereas in the
interpolated case they were all assigned to the low-mass branch of the
solution. This is not odd in itself.

Thus we conclude that consistent use of the isochrones throughout the
super-AGB and low-mass red supergiants regime leads to erroneous results.
Thereby validating our interpolation of the Mass--Luminosity and
Mass--Pulsation relations assuming that those stars do become cooler and more
luminous (cf.\ Ritossa, Garc\'{\i}a-Berro \& Iben 1996; Siess 2010) and that
they do pulsate over longer durations than suggested by the isochrones (cf.\
Ventura \& D'Antona 2011). As a consequence, we expect super-AGB stars to
develop strong, dusty winds (following the mass-loss prescription of van Loon
et al.\ 2005). This lends support to the suggestion by Botticella et al.\
(2009) that the dust-enshrouded progenitor of SN\,2008S was a super-AGB star.

We remind the reader that while the super-AGB stars undergo HBB and thus reach
higher luminosities than calculated in the isochrones, red supergiants do not
experience HBB and it is therefore not obvious that their evolution was also
foreshortened in the calculation of the isochrones. The exact mass range of
super-AGB stars is uncertain and could extend up to as high as $\sim11$
M$_\odot$ if convective overshoot is ineffective (Eldridge \& Tout 2004).

\subsection{Spatial variations}

Up till now, we have considered the SFH derived from the entire central square
kpc ($4^\prime\times4^\prime$) of M\,33. However, this region may well show
different galactic structures associated with populations of stars formed at
different times. Figure 1 in Paper I clearly shows a central concentration of
stars giving the appearance of a ``bulge'' of $\sim1'$ diameter. To examine
this, first we define a circular region of radius $R=110^{\prime\prime}$, to
avoid problems caused by the corners of the rectangular area covered in our
UIST monitoring survey. Secondly, we separate the stars in our catalogue into
massive stars, AGB stars, and RGB stars, on the basis of K-band magnitude and
J--K colour criteria (see Fig.\ 1): we define a demarcation line between hot
massive stars and cooler, less-massive giant stars to run from
$(J-K,K)=(0.6,18)$ to $(J-K,K)=(0.9,15.6)$ mag, with massive stars those that
have J--K colours bluer than this (down to $K=19.5$ mag) or that have $K<15.6$
mag, and AGB stars and RGB stars those that have J--K colours redder than this
and that have $16<K<18$ mag (AGB) or $18.3<K<19.5$ mag (RGB, which does extend
to much fainter magnitudes than our survey completeness). We left small gaps
to avoid contamination.

We consider two types of galactic structures: disc-related and spheroidal. The
former have axi-symmetry in the galaxy plane and are otherwise ``flat'' (with
relatively little depth), while the latter have a surface density that is
invariant to inclination with respect to the plane of the sky (as long as
internal extinction is negligible, which we have shown in Paper I is largely
the case for our near-IR survey). To investigate the disc-related structures,
we must deproject the positions of stars on the image to the galaxy plane --
but for the spheroidal structures we must make sure {\it not} to do that. To
deproject, we first rotate the image according to the position angle of
$PA=23^\circ$ and then stretch the coordinates along the direction of
inclination according to an inclination angle of $i=56^\circ$ (Zaritsky et
al.\ 1989).

%
%
\begin{figure}
\centerline{\psfig{figure=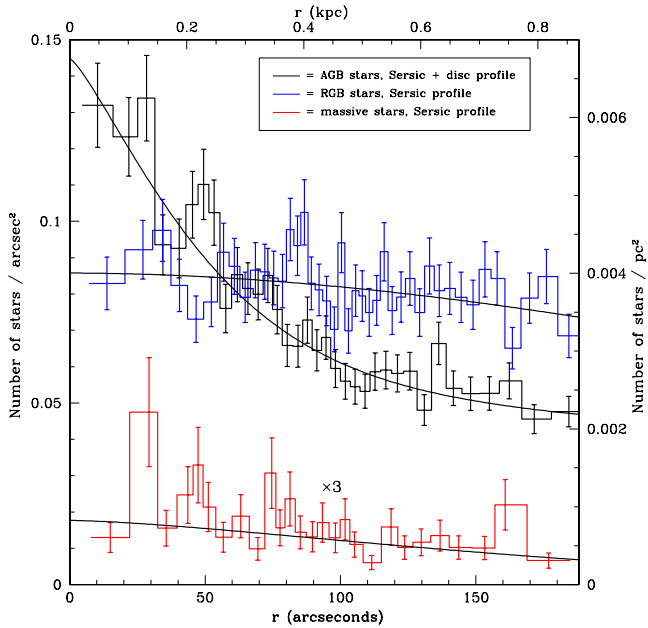,width=85mm}}
\caption[]{Radial distributions of massive stars (in red, multiplied by 3 for
ease of comparison), AGB stars (in black), and RGB stars (in blue), in the
galaxy plane. Each of these are fitted with a S\'ersic profile (the fit
parameters are listed in Table 6).}
\end{figure}

The radial distributions of stars as measured in the galaxy plane are shown in
Fig.\ 14. These are fitted with an adaptation for the stellar number density
of the surface brightness function from S\'ersic (1963):
\begin{equation}
N(r) = N_0 \exp\left(-b_n(r/R_{\rm e})^{1/n}-1\right),
\end{equation}
where $n=4$ returns the de Vaucouleurs' profile typical of galaxy bulges (de
Vaucouleurs 1953) and $n=1$ an exponential disc. The calibration of the
S\'ersic profile is such that half the intensity (number of stars in our case)
falls within $r=R_{\rm e}$. The parameter $b_n$ can be derived exactly from
comparison of the complete and incomplete Gamma functions (Ciotti 1991), and
approximately from $b_n=1.9992n-0.3271$ for $0.5<n<10$ (Capaccioli 1989). The
fit parameters are listed in Table 6.

The AGB distribution shows clear signs of a double-component profile, with the
break occurring around $r\sim0.4$ kpc, so in this case we fitted a S\'ersic
profile, with $R_{\rm e}=0.30$ kpc, plus a pure exponential disc profile:
\begin{equation}
N(r) = N_0 \exp\left(-r/R_{\rm e}\right).
\end{equation}
This corroborates the result found by Kormendy \& Kennicutt (2004) for the
surface brightness profile: a S\'ersic profile with $n=1.09\pm0.18$ and
$R_{\rm e}=0.31\pm0.05$ dominating the inner part with the outer part
described by an exponential disc. Clearly, the AGB stars do not constitute a
``bulge'', which would have had $n$ closer to 4. Apart from the pure
(rotating) disc component the nuclear part could possibly resemble a
pseudo-bulge (which is a disc phenomenon and as such has $n$ also close to 1).

The RGB population shows no strong central condensation, and a large effective
radius $R_{\rm e}\sim2$ kpc. This is easy to understand as the RGB stars have
had many Gyr to relax within the overall gravitational potential of the entire
M\,33 galaxy. The massive star population too, while more centrally
concentrated than the RGB population, is spread throughout the M\,33 disc,
without a strong nuclear component. This suggests that steady star formation
takes place throughout the M\,33 disc, with only a marginal concentration
towards the nucleus. Possibly, the epochs of enhanced star formation {\it are}
a feature of the central regions of M\,33, leading to the lasting imprint in
the AGB stars' distribution in the form of a pseudo-bulge.

%
%
\begin{figure}
\centerline{\psfig{figure=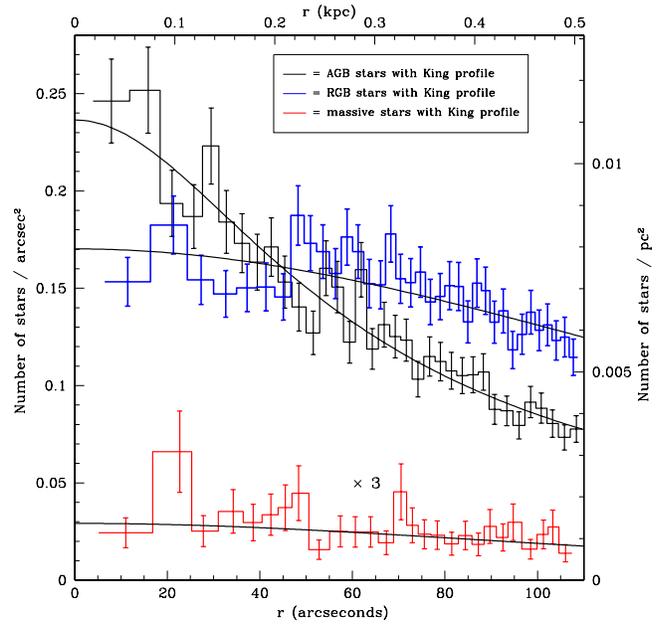,width=85mm}}
\caption[]{Radial distributions of massive stars (in red, multiplied by 3 for
ease of comparison), AGB stars (in black), and RGB stars (in blue), in the
image plane. Each of these are fitted with a King profile (the fit parameters
are listed in Table 6).}
\end{figure}

The radial distributions of stars as measured in the image plane are shown in
Fig.\ 15. These are fitted with one of the various King models (King 1962):
\begin{equation}
N(r) = N_0\left(1+\left(r/R_{\rm c}\right)^2\right)^{-\frac{3}{2}\beta}.
\end{equation}
The fit parameters are listed in Table 6.

Interestingly, the fits are not significantly better or worse than for the
S\'ersic profiles fitted in the galaxy plane. While it may seem reasonable to
expect that the old RGB stars constitute a pressure-supported, more spheroidal
system, the success of the S\'ersic fit with a small value for $n$ suggests
that perhaps they form part of a more flattened system. In either case the RGB
population has the largest characteristic radius among the three populations.
The massive stars and at least a fraction of the intermediate-age AGB stars
are not expected to reside in a spheroidal configuration, as they form within
the gas-rich disc and will not have had time to reconfigure into a dynamically
relaxed system -- for these populations the S\'ersic profile fits yield more
plausible results.

%
%
\begin{table}
\caption{Fit parameters of S\'ersic and King profiles for massive stars, AGB
stars (S\'ersic+disc), and RGB stars in the central square kpc of M\,33.}
\begin{tabular}{lccccc}
\hline\hline
\multicolumn{6}{c}{\it S\'ersic profile} \\
Population
 & \multicolumn{2}{c}{$N_0$}     & \multicolumn{2}{c}{$R_{\rm e}$} & $n$ \\
 & (arcsec$^{-2}$) & (pc$^{-2}$) & ($^{\prime\prime}$) & (kpc)     &     \\
\hline
Massive stars & 0.002\rlap{1} & 0.0001\rlap{0} & \llap{1}99 & 0.92 & 0.67 \\
AGB stars     & 0.025         & 0.0012         &         64 & 0.30 & 0.78 \\
(exp.\ disc:) & 0.050         & 0.0023         & \llap{8}46 & 3.92 &      \\
RGB stars     & 0.042         & 0.0020         & \llap{4}20 & 1.94 & 0.52 \\
\hline
\multicolumn{6}{c}{\it King profile} \\
Population
 & \multicolumn{2}{c}{$N_0$}     & \multicolumn{2}{c}{$R_{\rm c}$} & $\beta$ \\
 & (arcsec$^{-2}$) & (pc$^{-2}$) & ($^{\prime\prime}$) & (kpc)     &         \\
\hline
Massive stars & 0.01\rlap{0} & 0.000\rlap{5} & \llap{1}72 & 0.80 & 1   \\
AGB stars     & 0.24         & 0.011         &         47 & 0.22 & 0.4 \\
RGB stars     & 0.17         & 0.008         & \llap{2}29 & 1.06 & 1   \\
\hline
\end{tabular}
\end{table}

%
%
\begin{figure}
\centerline{\psfig{figure=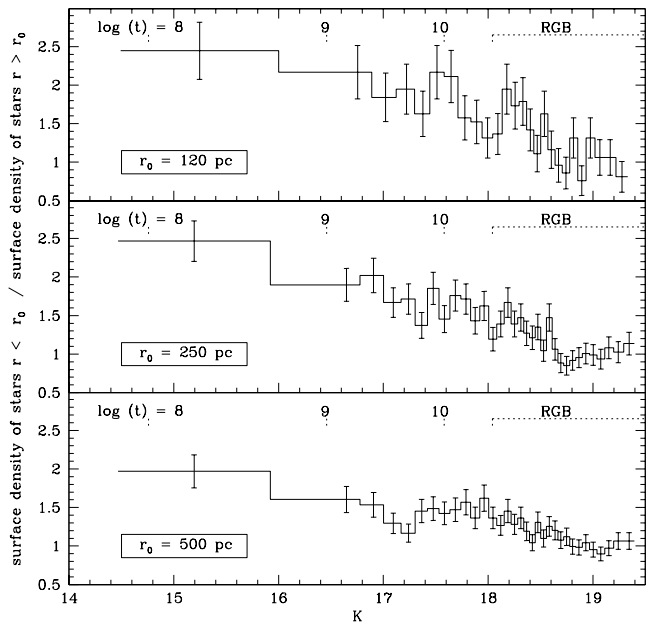,width=85mm}}
\caption[]{Ratio of stellar density inside and outside of 120 pc (top panel),
250 pc (middle panel), or 500 pc (bottom panel) for the massive stars, AGB
stars, and RGB stars. While the RGB populations are not centrally concentrated
(a ratio near unity) the AGB and massive stellar populations are progressively
more centrally concentrated the younger they are.}
\end{figure}

Another way to illustrate the central condensation as a function of population
age is achieved by comparing the average stellar surface density within and
outside a given radius (Fig.\ 16; the radius is defined in the galaxy plane).
This suggests that star formation has gradually become more concentrated
towards the central regions, or that star formation has always been
concentrated towards the core with dynamical relaxation having dispersed stars
as time went by. We stress that we are looking here at variations within the
inner square kpc, and that further out the situation might well differ.

%
%
\begin{figure}
\centerline{\psfig{figure=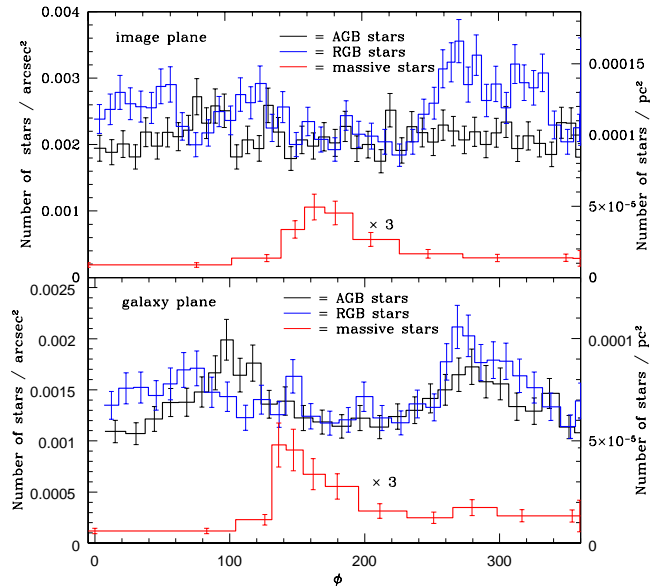,width=85mm}}
\caption[]{Azimuthal distributions of the massive stars (in red), AGB stars
(in black), and RGB stars (in blue), in the image plane (top panel) and galaxy
plane (bottom panel).}
\end{figure}

One could wonder whether we are already noticing the gravitational potential
responsible for the spiral arm pattern that so clearly defines M\,33. Or,
would the pseudo-bulge be tri-axial, like a bar potential? In either case we
would expect an azimuthal modulation with two maxima and two minima offset by
an angle $\Delta\phi=\pi$. In Fig.\ 17 we plot these distributions for the
three stellar populations, both in the image plane as well as in the galaxy
plane. These distributions are constructed only for stars within circular
areas in the image and galaxy planes, respectively, so as not to introduce
spurious modulations due to the corners of the image resulting from the
generally falling stellar surface density with distance to the galaxy centre.

The AGB population seems not to show any significant azimuthal structure in
the image plane, but in the galaxy plane it does show the modulation expected
from a bar or two-armed spiral arm pattern, with stellar surface density peaks
around $\phi_{\rm galaxy}\sim100^\circ$ and $280^\circ$, i.e.\ roughly in the
East and West directions. The RGB population shows a similar distribution in
the galaxy plane, with an enhancement also visible in the image plane around
$\phi_{\rm image}\sim260$--$330^\circ$, i.e.\ Westwards. The massive stars, on
the other hand, show a clear overdensity towards the South, $\phi_{\rm
image}\sim170^\circ$.

\section{Discussion and conclusions}

%
%
\begin{figure*}
\centerline{\hbox{
\psfig{figure=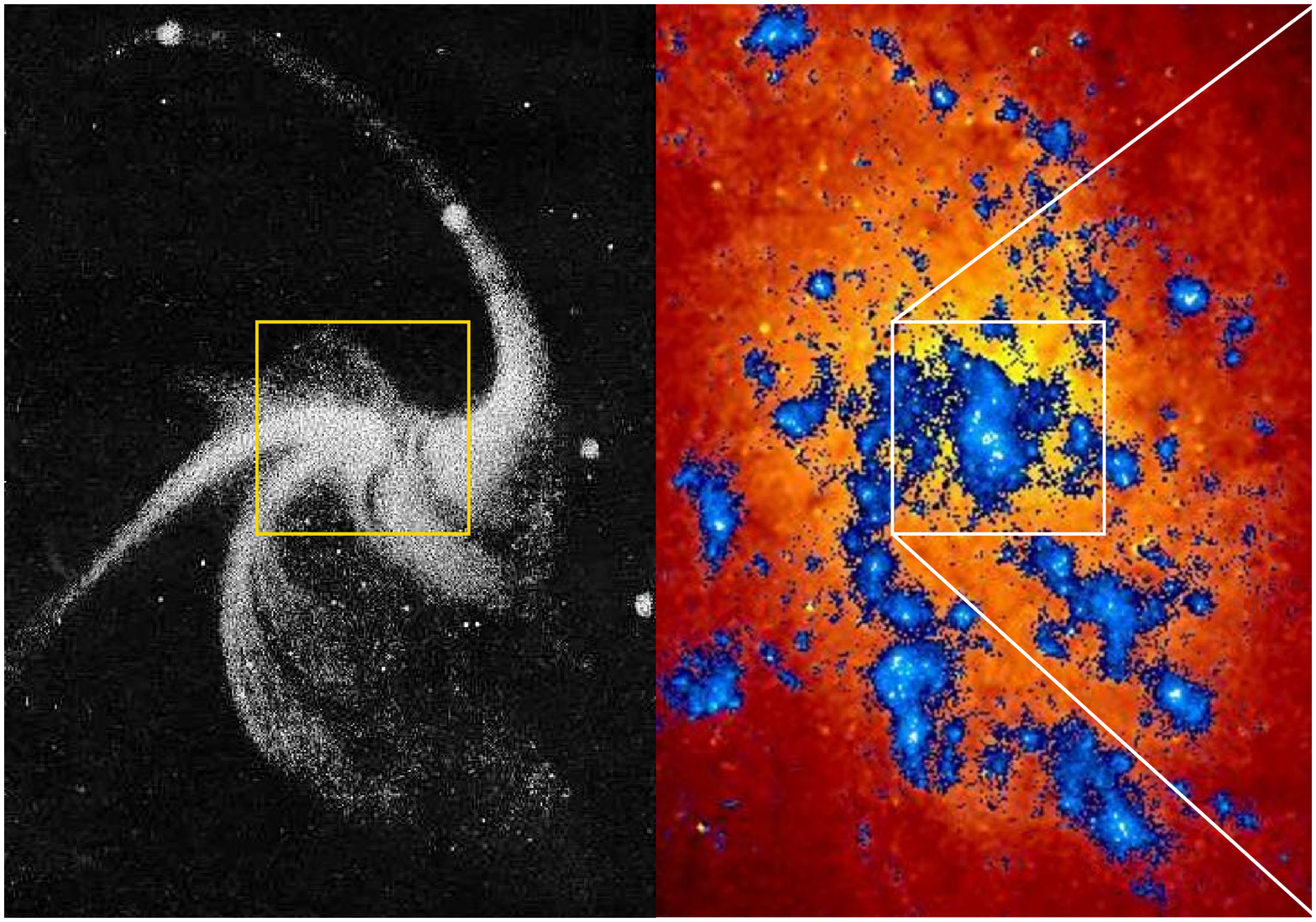,width=103.5mm}
\psfig{figure=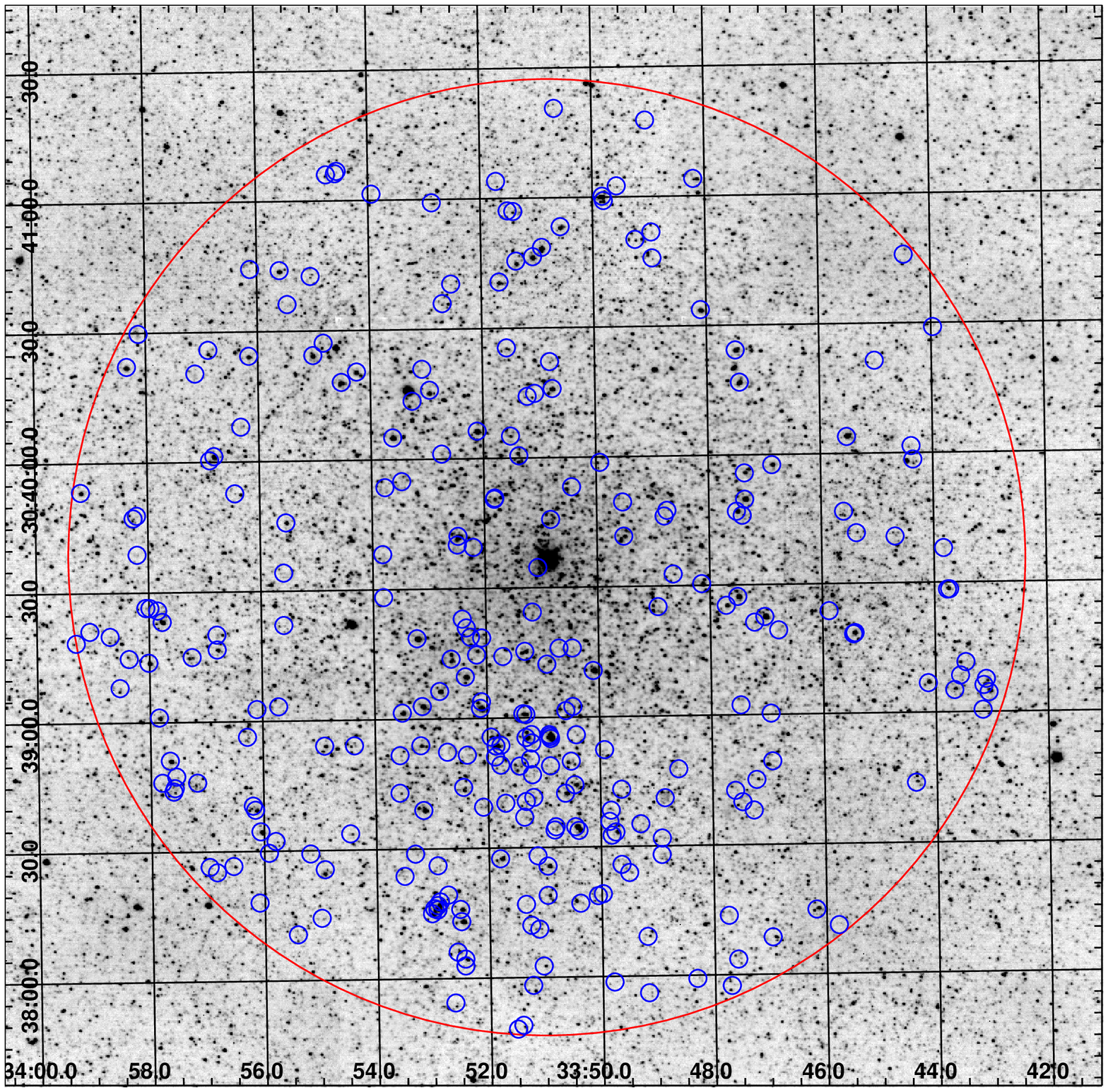,width=73.3mm}
}}
\caption[]{(Left:) Drawing of M\,33 made by R.J.\ Mitchell (Rosse 1850).
(Middle:) Ultraviolet Imaging Telescope image in blue, against an optical
image in yellow/red (Landsman et al.\ 1992). (Right:) UFTI image of the
central square kpc of M\,33 (the box in the left panels), with overlain the
massive stars (circles) and the circular area within which we constructed the
azimuthal distribution in the image plane.}
\end{figure*}

The photometric catalogue of Javadi et al.\ (2011a) was used to reconstruct
the star formation history and structure of the inner square kpc of M\,33. The
numbers and luminosities of the pulsating AGB stars and red supergiants were
converted to star formation rates as a function of look-back time, using
recent Padova stellar evolution models (Marigo et al.\ 2008). Consistency
checks between the predicted luminosities and pulsation durations and the
derived star formation rates lend strong support for these models except that
the super-AGB stars must also reach high luminosities and pulsate for a
considerable amount of time to avoid implausible values for the inferred star
formation rates. This also supports the interpretation of dust-enshrouded
supernovae as being due to super-AGB star progenitors.

Our conclusions pertaining to the star formation history in the centre of
M\,33 can be summarised as follows:
\begin{itemize}
\item[$\bullet$]{The disc of M\,33 was built $\geq6$ Gyr ago ($\log
t\geq9.8$), when $\geq80$\% of the historic star formation in M\,33 occurred.}
\item[$\bullet$]{Shorter epochs of enhanced star formation (may) have occurred
since then; the most recent substantial example of such event occurred
$\sim250$ Myr ago ($\log t\sim8.4$) and contributed $\leq6$\% to M\,33's
historic star formation.}
\item[$\bullet$]{A minor star formation event which had been suggested to have
occurred within the nuclear star cluster within the past $\sim40$ Myr ($\log
t\lsim 7.6$) can also be discerned in our data suggesting this was a more
wide-spread phenomenon.}
\end{itemize}

The formation epoch of M\,33 around $t\geq6$ Gyr is in line with the peak in
the distribution of formation epochs of bulgeless galaxies in a standard
$\Lambda$-CDM cosmological model, around redshift $z\sim1$ (Fontanot et al.\
2011) corresponding to ages $t\sim7$ Gyr. We speculate that more recent star
formation epochs may have been induced by interaction with the neighbouring
more massive Andromeda spiral galaxy (M\,31; Richardson et al.\ 2011).

Comparison with the historical drawing of M\,33 by R.J.\ Mitchell and
published in Lord Rosse's records shows a reasonable correspondence between
the footpoints of the spiral-arm pattern revealed in the unresolved white
light and the density enhancements that we recorded (Fig.\ 18). The two main
spiral arms seem to originate around $\phi_{\rm image}\sim90$--$120^\circ$ and
$\sim270^\circ$ with a Southern spur visible close to the centre. The spiral
arms may be associated with the overdensities in AGB stars and RGB stars,
though it would be somewhat coincidental that the signal in the AGB
distribution is so perfectly cancelled by the projection on the sky. Also, one
might not have expected a signal this clear in the distribution of RGB stars,
if these occupy a more spheroidal, dynamically-relaxed distribution. Possibly
we are dealing with a bar-like feature, which is a disc-related structure and
may be connected to the footpoints of the spiral arms -- in fact the S\'ersic
index $n\sim0.5$--0.8 is smaller than that of a pure exponential disc, with
galactic bars generally having $n\sim0.5$.

The Southern spur, on the other hand, can be associated with the overdensity
of massive stars, which are displayed on our UFTI image (Fig.\ 18). This
includes a prominent cluster at $(RA,Dec)\sim(1^{\rm h}33^{\rm m}53^{\rm
s},+30^\circ38^\prime15^{\prime\prime}$). The Southern spur is also prominent
in ultraviolet images obtained with the Ultraviolet Imaging Telescope on-board
space shuttle Columbia (Landsman et al.\ 1992). Note also that the region
within $\sim30^{\prime\prime}$ is actually rather devoid of massive stars;
this is also where the pseudo-bulge (or bar) is seen, which therefore seems to
be a predominantly intermediate-age feature.

\section*{Acknowledgments}

We thank the staff at UKIRT for their excellent support of this programme.
Discussions with L\'eo Girardi, Paola Marigo and Lionel Siess helped us decide
on how to treat the super-AGB stars, and Dean McLaughlin kindly shared some of
his knowledge about galactic structure. AJ wishes to thank Dr.\ Habib
Khosroshahi for valuable advice at various stages of the project. We also
thank the anonymous referee for her/his positive report which helped us to
improve the presentation of our work. We are grateful for financial support by
The Leverhulme Trust under grant No.\ RF/4/RFG/2007/0297.


\label{lastpage}

\end{document}